\documentclass[a4paper,11pt]{article}
\pdfoutput=1 
\usepackage{jheppub} 
\usepackage[T1]{fontenc} 
\RequirePackage[displaymath]{lineno} 

\usepackage{epsfig}
\usepackage{graphicx}
\usepackage{dcolumn}
\usepackage{bm}
\usepackage{ltablex,booktabs}
\usepackage{overpic}
\usepackage{subfigure}
\usepackage{float}
\usepackage{color}
\usepackage{amsmath}
\usepackage{mathcomp}
\usepackage{mathrsfs}
\usepackage{multirow}
\usepackage{rotating}
\usepackage{amssymb}
\usepackage{gensymb}
\usepackage{amsmath}
\usepackage{tabularx}
\usepackage{epstopdf}

\title{Amplitude analysis and branching fraction measurement of \boldmath $D_{s}^{+} \to K^-K^+\pi^+\pi^+\pi^-$}
\collaborationImg{\includegraphics[width=0.15\textwidth, angle=90]{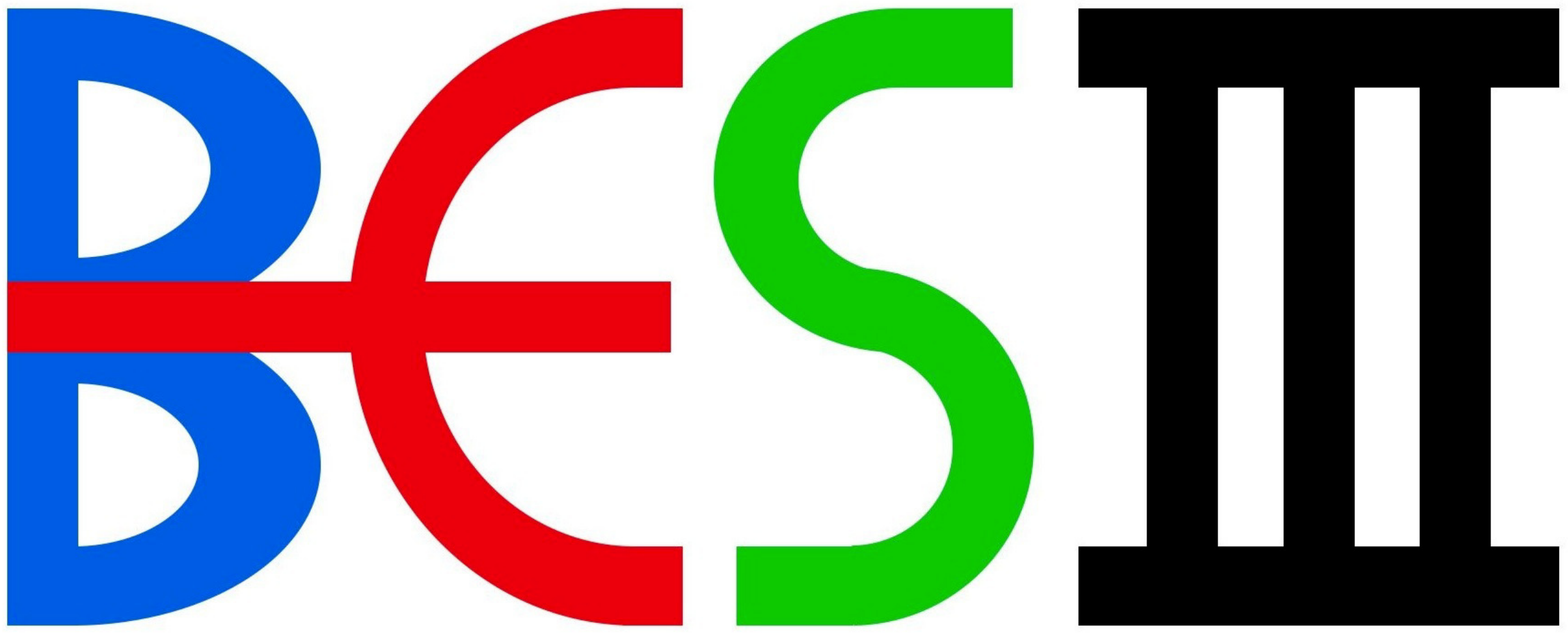}}
\collaboration{The BESIII Collaboration}

\date{\today}
\abstract{Using $e^{+}e^{-}$ annihilation data corresponding to a total integrated luminosity of $6.32$ fb$^{-1}$ collected at the center-of-mass energies between 4.178 and 4.226 GeV with the BESIII detector, 
we perform an amplitude analysis of the decay $D^+_s \to K^-K^+\pi^+\pi^+\pi^-$ and determine the relative fractions and phases of different intermediate processes.
Absolute branching fraction of $D^+_s\to K^-K^+\pi^+\pi^+\pi^-$ decay is measured to be ($6.60\pm0.47_{\rm stat.}\pm0.35_{\rm syst.})\times 10^{-3}$.
The dominant intermediate process is $D_{s}^{+} \to a_1(1260)^+\phi, \phi\to K^-K^+, a_1(1260)^+\to \rho\pi^+, \rho\to\pi^+\pi^-$, with a branching fraction of $(5.16\pm0.41_{\rm stat.}\pm0.27_{\rm syst.})\times 10^{-3}$.}

\keywords{BESIII, $D_s$ meson, amplitude analysis, five-body decay}
\arxivnumber{}
\begin{document}
\maketitle
\flushbottom


\section{Introduction}
\begin{figure}[htbp]
  \centering
  \includegraphics[width=0.6\textwidth]{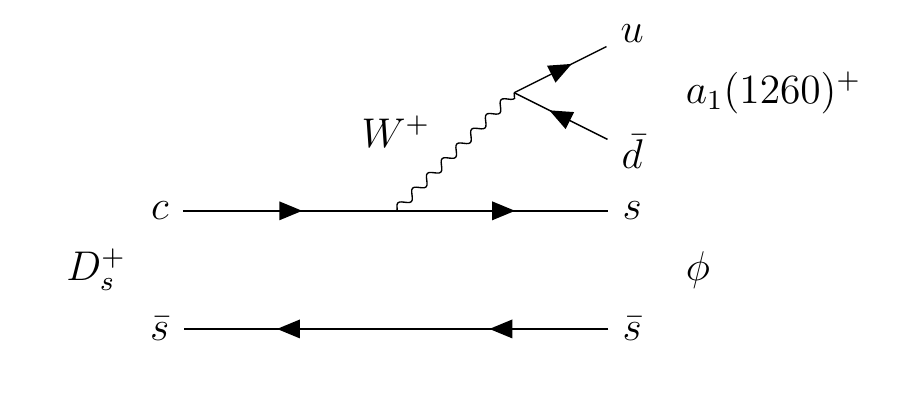}
  \caption{Feynman diagram of $D^+_s\to a_1(1260)^+ \phi$.}
  \label{fig:feynman}
\end{figure}

The hadronic decays of $D^+_s$ mesons are dominated by two-body processes~\cite{PDG}, such as $D_s^+ \to PP$, $VP$, $VS$, $VV$, $AP$ and $AV$, where $P$, $V$, $S$ and $A$ denote pseudo-scalar, vector, scalar and axial-vector mesons, respectively.
The branching fractions (BFs) of most of these decays can be calculated theoretically~\cite{Lipkin:2000gz}, even if the non-perturbative contributions, such as final-state interactions, make some of them hard to predict.
Therefore, BFs measurements of the $D_s^+$ two-body decays are important to test the theoretical calculations and can be helpful to understand the decay mechanisms of $D_s^+$ mesons.
Up to now, there are no references to studies focusing on $D_s^+ \to AV$ decays.
Among them, the process $D^+_s \to a_1(1260)^+\phi$, which is mediated via the diagram in Fig.~\ref{fig:feynman}, can be studied in the $D_s^+ \to K^-K^+\pi^+\pi^+\pi^-$ decay.

Moreover, experimental study of $D^+_s\to K^+K^-\pi^+\pi^+\pi^-$ is also helpful to clarify the tension observed in the ratio $R(D^{*}) \equiv \mathcal{B}(B\to D^{*}\tau^+\nu_{\tau})/\mathcal{B}(B\to D^{*}\ell^+\nu_{\ell})$ $(\ell = e, \mu)$, 
with an average of $0.295\pm0.011\pm0.008$ provided by the Heavy Flavor Averaging Group, which differs from the Standard Model prediction of $0.258\pm0.005$ by 2.6 standard deviations~\cite{HFLAV:2016hnz}.
This hints a possibile violation of the lepton flavor universality. 
However, the $R(D^*)$ measurement at LHCb experiment suffers from a large systematic uncertainty due to the limited knowledge of the inclusive $D^+_s\to \pi^+\pi^+\pi^-X$ decay~\cite{LHCb:2017smo, LHCb:2017rln}, where the decay of $B\to D^{*-}D^+_s, D^+_s\to \pi^+\pi^+\pi^-$X is main background in the decay chain $B^0\to D^{*-}\tau^+\nu_{\tau}, \tau^+\to \pi^+\pi^+\pi^-$X.
A precise measurement of the branching fraction~(BF) of $D^+_s\to K^-K^+\pi^+\pi^+\pi^-$, which is one of the dominant processes in $D^+_s\to\pi^+\pi^+\pi^-X$, can provide a useful input to improve the precision of $R(D^*)$.

The E687~\cite{E687:1997dng} and FOCUS~\cite{FOCUS:2002psb} experiments reported the fitted yields and BFs of $D^+_s\to K^-K^+\pi^+\pi^+\pi^-$ relative to $D^+_s\to K^+K^-\pi^+$, as shown in Table~\ref{E687}. 
By performing an analysis of the resonant substructure in the decay $D^+_s\to K^-K^+\pi^+\pi^+\pi^-$, the FOCUS experiment observed that the decay proceeds primarily through a quasi-two-body decay involving an $a_1(1260)^+$ resonance.
In this paper, we perform the first amplitude analysis, as well as measuring the absolute BF of $D^+_s\to K^-K^+\pi^+\pi^+\pi^-$, by using the 6.32 fb$^{-1}$ data samples collected by the BESIII detector at the center-of-mass energies ($\sqrt{s}$) from 4.178 to 4.226 GeV.
More precise measurements and a detailed study of the decay structure are expected. 
Charge conjugate states are always implied throughout this paper.

\begin{table}[htb]
	\centering
	\small
	\caption{The fitted yields and BF ratios for the previous measurements by E687 and FOCUS experiments. All BF ratios are inclusive of subresonant modes.}
	\begin{tabular}{c c ccc}
		\hline
		 &\multicolumn{2}{c}{E687~\cite{E687:1997dng}}  &\multicolumn{2}{c}{FOCUS~\cite{FOCUS:2002psb}} \\
    \hline
		Decay mode &fitted yield &BF ratio &fitted yield &BF ratio\\
		\hline
		$\frac{\Gamma(D^+_s\to K^-K^+\pi^+\pi^+\pi^-)}{\Gamma(D^+_s\to K^-K^+\pi^+)}$ &$240\pm30$ &$0.188\pm0.036\pm0.040$ &$136\pm14$ &$0.150\pm0.019\pm0.025$ \\
		$\frac{\Gamma(D^+_s\to \phi\pi^+\pi^+\pi^-)}{\Gamma(D^+_s\to K^-K^+\pi^+)}$   &$75\pm13$ &$0.280\pm0.060\pm0.010$ &$40\pm8$ &$0.249\pm0.024\pm0.021$ \\
    \hline
		\end{tabular}
		\label{E687}
\end{table}

\section{Detector and data sets}
\label{sec:detector_dataset}
The BESIII detector~\cite{Ablikim:2009aa} records symmetric $e^+e^-$ collisions provided by the BEPCII storage ring~\cite{Yu:IPAC2016-TUYA01}, 
which operates in the center-of-mass energy range from $\sqrt{s}=2.00$ to $\sqrt{s}=4.95$~GeV~\cite{Ablikim:2019hff}. 
The cylindrical core of the BESIII detector covers 93\% of the full solid angle and consists of a helium-based multilayer drift chamber~(MDC), 
a plastic scintillator time-of-flight system~(TOF), and a CsI(Tl) electromagnetic calorimeter~(EMC), 
which are all enclosed in a superconducting solenoidal magnet providing a 1.0~T magnetic field. 
The solenoid is supported by an octagonal flux-return yoke with resistive plate counter muon identification modules interleaved with steel.
The charged-particle momentum resolution at $1~{\rm GeV}/c$ is $0.5\%$, and the $dE/dx$ resolution is $6\%$ for electrons from Bhabha scattering. 
The EMC measures photon energies with a resolution of $2.5\%$ ($5\%$) at $1$~GeV in the barrel (end cap) region. 
The time resolution in the TOF barrel region is 68~ps, while that in the end cap region is 110~ps. 
The end cap TOF system was upgraded in 2015 using multi-gap resistive plate chamber technology, providing a time resolution of 60~ps~\cite{etof1, etof2, etof3}.

The data samples used in this analysis contain a total integrated luminosity of 6.32~fb$^{-1}$ collected at the center-of-mass energies between 4.178 and 4.226~GeV. 
The integrated luminosity of each data sample is shown in Table~\ref{energe}~\cite{luminosities}. 
The data sets are organized into three sample groups, 4.178~GeV, 4.189-4.219~GeV, and 4.226~GeV, which were acquired during the same year under consistent running conditions.

 \begin{table}[htb]
 \renewcommand\arraystretch{1.25}
 \centering
 \caption{The integrated luminosities ($\mathcal{L}_{\rm int}$) and the requirements on $M_{\rm rec}$ for various center-of-mass energies. 
	 The definition of $M_{\rm rec}$ is given in Eq.~(\ref{eq:mrec}). 
	 The first and second uncertainties are statistical and systematic, respectively.}
 \begin{tabular}{ccc}
 \hline
 $\sqrt{s}$ (GeV) & $\mathcal{L}_{\rm int}$ (pb$^{-1}$)~\cite{luminosities} & $M_{\rm rec}$ (GeV/$c^2$)\\
 \hline
  4.178 &3189.0$\pm$0.2$\pm$31.9&[2.050, 2.180] \\
  4.189 &526.7$\pm$0.1$\pm$2.2&[2.048, 2.190] \\
  4.199 &526.0$\pm$0.1$\pm$2.1&[2.046, 2.200] \\
  4.209 &517.1$\pm$0.1$\pm$1.8&[2.044, 2.210] \\
  4.219 &514.6$\pm$0.1$\pm$1.8&[2.042, 2.220] \\
  4.226 &1056.4$\pm$0.1$\pm7.0$&[2.040, 2.220] \\
  \hline
 \end{tabular}
 \label{energe}
\end{table}
Inclusive Monte Carlo (MC) samples that are 40 times larger than the data sets are produced at the center-of-mass energies between 4.178 and 4.226 GeV with a {\sc geant4}-based~\cite{GEANT4:2002zbu} MC package, 
which includes the geometric description of the BESIII detector and the detector response. These samples are used to determine detection efficiencies and to estimate backgrounds.
The samples include the production of open charm processes, the initial-state radiation (ISR) production of vector charmonium(-like) states and the continuum processes incorporated in {\sc kkmc}~\cite{Jadach:2000ir, Jadach:1999vf}. 
The known decay modes are modeled with {\sc evtgen}~\cite{Lange:2001uf, EVTGEN2} using the BFs taken from the Particle Data Group (PDG)~\cite{PDG}, and the remaining unknown charmonium decays are modeled with {\sc lundcharm}~\cite{Chen:2000tv, LUNDCHARM2}. 
Final state radiation~(FSR) from charged final state particles is incorporated using {\sc photos}~\cite{PHOTOS}. 
A phase-space (PHSP) MC sample is produced with the $D_s^+ \to K^-K^+\pi^+\pi^+\pi^-$ generated with a uniform distribution and is used to extract the detection efficiency maps.
Initially, the PHSP MC sample is used to calculate the normalization integral used in the determination of the amplitude model parameters in the fit to data. 
Then, the signal MC sample is re-generated with the $D_s^+$ meson decaying to $K^-K^+\pi^+\pi^+\pi^-$ using the amplitude model. 
It is used to evaluate the fit quality and estimate the systematic uncertainty.

\section{Event selection}
\label{ST-selection}
The data samples were collected just above the $D_s^{*\pm}D_s^{\mp}$ threshold. 
In this energy region, pairs of $D_s^{*\pm}D_s^{\mp}$ mesons are produced copiously; subsequently, the $D_s^{*\pm}$ meson predominantly decays to $\gamma D_s^{\pm}$ with a branching fraction of ($93.5\pm0.7$)$\%$~\cite{PDG}. 
The tag method~\cite{MARK-III:1985hbd} allows to select clean signal samples, providing the opportunity to perform amplitude analyses and to measure the absolute BFs of the hadronic $D^+_s$ meson decays. 
In the tag method, a single-tag (ST) candidate requires only one of the $D_{s}^{\pm}$ mesons to be reconstructed via a hadronic decay. 
A double-tag (DT) candidate has both $D_s^+D_s^-$ mesons reconstructed, requiring the $D_{s}^{+}$ meson decaying to the signal mode $D_{s}^{+} \to K^-K^+\pi^{+}\pi^{+}\pi^{-}$ and the $D_{s}^{-}$ meson decaying to the eight tag modes listed in Table~\ref{tab:tag-cut}. 
The reconstruction of $\pi^\pm$, $K^\pm$, $K^0_S$, $\pi^0$, $\eta$, and $\eta^\prime$ particles in the final state is discussed below.

\begin{table}[http]
 \renewcommand\arraystretch{1.25}
 \centering
 \caption{Requirements on $M_{\rm tag}$ for various tag modes, where the $\eta$
   and $\eta^\prime$ subscripts denote the decay modes used to reconstruct
   these particles.}\label{tab:tag-cut}
	 \label{tagwindow}
     \begin{tabular}{lc}
        \hline
        Tag mode                                     & Mass window (GeV/$c^{2}$) \\
        \hline
        $D_{s}^{-} \to K_{S}^{0}K^{-}$               & [1.948, 1.991]            \\
        $D_{s}^{-} \to K^{+}K^{-}\pi^{-}$            & [1.950, 1.986]            \\
        $D_{s}^{-} \to K_{S}^{0}K^{+}\pi^{0}$        & [1.946, 1.987]            \\
        $D_{s}^{-} \to K^{+}K^{-}\pi^{-}\pi^{0}$     & [1.947, 1.982]            \\
				$D_{s}^{-} \to K_{S}^{0}K^{-}\pi^{-}\pi^{+}$ & [1.958, 1.980]            \\
        $D_{s}^{-} \to K_{S}^{0}K^{+}\pi^{-}\pi^{-}$ & [1.953, 1.983]            \\
        $D_{s}^{-} \to \pi^{-}\eta_{\pi^{+}\pi^{-}\eta_{\gamma\gamma}}^{\prime}$ 
				                                             & [1.940, 1.996]            \\
				$D_{s}^{-} \to \pi^{-}\eta_{\gamma\gamma}$   & [1.930, 2.000]            \\
        \hline
      \end{tabular}
\end{table}

All charged tracks reconstructed in the MDC must satisfy the condition $|$cos$\theta|<0.93$, where $\theta$ is the polar angle with respect to the direction of the positron beam. 
For charged tracks not originating from $K_S^0$ decays, the distance of closest approach to the interaction point is required to be less than 10~cm along the beam direction and less than 1~cm in the plane perpendicular to the beam.
Particle identification~(PID) for charged tracks combines the measurements of the $dE/dx$ in the MDC and the time of flight in the TOF to form likelihoods $\mathcal{L}(h)~(h=K,\pi)$ for each hadron $h$ hypothesis.
Charged kaons and pions are identified by comparing the likelihoods for the two hypotheses, $\mathcal{L}(K)>\mathcal{L}(\pi)$ and $\mathcal{L}(\pi)>\mathcal{L}(K)$, respectively.

The $K_{S}^0$ candidates are selected by looping over all pairs of tracks with opposite charges, whose distances to the interaction point along the beam direction are within $20$ cm.
A primary vertex and a secondary vertex are reconstructed and the decay length between the two vertexes is required to be greater than twice its uncertainty. 
This requirement is not applied for the $D_s^-\to K_S^0K^-$ decay since the combinatorial background is low. 
Candidate $K_{S}^0$ particles are required to have the $\chi^2$ of the vertex fit less than 100 and an invariant mass of the $\pi^{+}\pi^{-}$ pair ($M_{\pi^{+}\pi^{-}}$) in the range $[0.487, 0.511]$ GeV$/c^{2}$. 
To prevent an event being double counted in the $D_s^-\to K_S^0K^-$ and $D_s^-\to K^-\pi^{+}\pi^{-}$ selections, the value of $M_{\pi^{+}\pi^{-}}$ is required to be outside of the mass range $[0.487, 0.511]$ GeV$/c^{2}$ for the $D_s^-\to K^-\pi^{+}\pi^{-}$ decay.

Photon candidates are identified using showers in the EMC. 
The deposited energy of each shower must be more than 25~MeV in the barrel region~($|\!\cos \theta|< 0.80$) and more than 50~MeV in the end cap region~($0.86 <|\!\cos \theta|< 0.92$). 
The angle between the position of each shower in the EMC and the closest extrapolated charged track must be greater than 10 degrees to exclude showers that originate from charged tracks. 
The difference between the EMC time and the event start time is required to be within [0, 700]\,ns to suppress electronic noise and showers unrelated to the event.

The $\pi^0$ $(\eta)$ candidates are reconstructed through $\pi^0\to \gamma\gamma$ ($\eta \to \gamma\gamma$) decays, with at least one photon falling in the barrel region. 
The invariant mass of the photon pair for $\pi^{0}$ and $\eta$ candidates must be in the ranges $[0.115, 0.150]$~GeV/$c^{2}$ and $[0.500, 0.570]$~GeV/$c^{2}$, respectively, which are about three times the mass resolution around their known masses. 
A kinematic fit that constrains the $\gamma\gamma$ invariant mass to the $\pi^{0}$ or $\eta$ known mass~\cite{PDG} is performed to improve the mass resolution and the $\chi^2$ is required to be less than 30. 
The $\eta^{\prime}$ candidates are formed from the $\pi^{+}\pi^{-}\eta$ combinations with an invariant mass within a range of $[0.946, 0.970]$~GeV/$c^{2}$.

Eight tag modes are reconstructed and the corresponding mass windows on the tagging $D_{s}^{-}$ mass~($M_{\rm tag}$) are listed in Table~\ref{tab:tag-cut}. 
The signal $D_{s}^{+}$ candidates, whose $M_{\rm rec}$ lies within the mass windows listed in Table~\ref{energe}, are retained for further studies, 
considering the quantity $M_{\rm rec}$ defined as
\begin{eqnarray}
\begin{aligned}
	\begin{array}{lr}
	M_{\rm rec} = \sqrt{\left(E_{\rm cm} - \sqrt{|\vec{p}_{D_{s}^-}|^{2}+m_{D_{s}^-}^{2}}\right)^{2} - |\vec{p}_{D_{s}^-} | ^{2}} \; , \label{eq:mrec}
		\end{array}
\end{aligned}\end{eqnarray}
where $E_{\rm cm}$ is the energy of the initial state calculated from the beam energy, 
$\vec{p}_{D_{s}^-}$ is the three-momentum of the $D_{s}^{-}$ candidate in the $e^+e^-$ center-of-mass frame, and $m_{D_{s}^-}$ is the $D_{s}^{-}$ known mass~\cite{PDG}.

\section{Amplitude analysis}
\label{Amplitude-Analysis}
\subsection{Further selection criteria}
\label{AASelection}
The following selection criteria are further applied in order to obtain signal samples with high purity for the amplitude analysis. 
The selection criteria discussed in this section are not used in the BF measurements.

A six-constraint (6C) kinematic fit is performed to the process $e^+e^-\to D^{*\pm}_s D^\mp_s \to \gamma D^+_s D^-_s$,
	assuming $D_s^-$ decaying to one of the tag modes and $D_s^+$ decaying to the signal mode ($K^- K^+ \pi^+\pi^+\pi^-$) with two hypotheses:
the signal $D_s^+$ comes from a $D^{*+}_s$ or the $D_s^-$ comes from a $D^{*-}_s$.
The total four-momentum is constrained to the initial four-momentum of the $e^+e^-$ system, and the invariant masses of tag $D^-_s$ and $D^{*\pm}_s$ candidates are constrained to the corresponding known masses. 
	The best $D^{*\pm}_sD^{\mp}_s$ combination with the minimum $\chi_{6C}^2$ is selected. 
Then, a seventh constraint of the signal $D_s^+$ invariant mass is added to the 6C kinematic fit, in order to ensure that all the events fall into the phase-space~(PHSP) boundary.
The updated four-momenta obtained from the seven-constraint kinematic fit of the final particles are used to perform the amplitude analysis.

The signal region of the $D_s^+$ invariant mass for the $D_s^+\to K^-K^+\pi^+\pi^+\pi^-$ decay is defined as $[1.955, 1.982]$~GeV/$c^2$. 
The $\pi^+$ with lower invariant mass of $\pi^+\pi^-$ pair is denoted as $\pi^+_1$ and the other $\pi^+$ as $\pi^+_2$.
If the invariant mass of $\pi^+_1\pi^-$ or $\pi^+_2\pi^-$ satisfies the selection of $K_S^0$ mesons, these candidates are vetoed. 
There are background events from $D^+_s\to K^-K^+\pi^+\pi^0, \pi^0\to e^+e^-\gamma$, in which the $e^+e^-$ pair is misidentified as a $\pi^+\pi^-$ pair. 
We use cos$\theta(\pi^+_1\pi^-)> 0.985$ to suppress this kind of background, where $\theta$ is the opening angle between the momenta of $\pi^-$ and $\pi^+_1$.

\begin{figure*}[!htbp]
  \centering
  \includegraphics[width=6.cm]{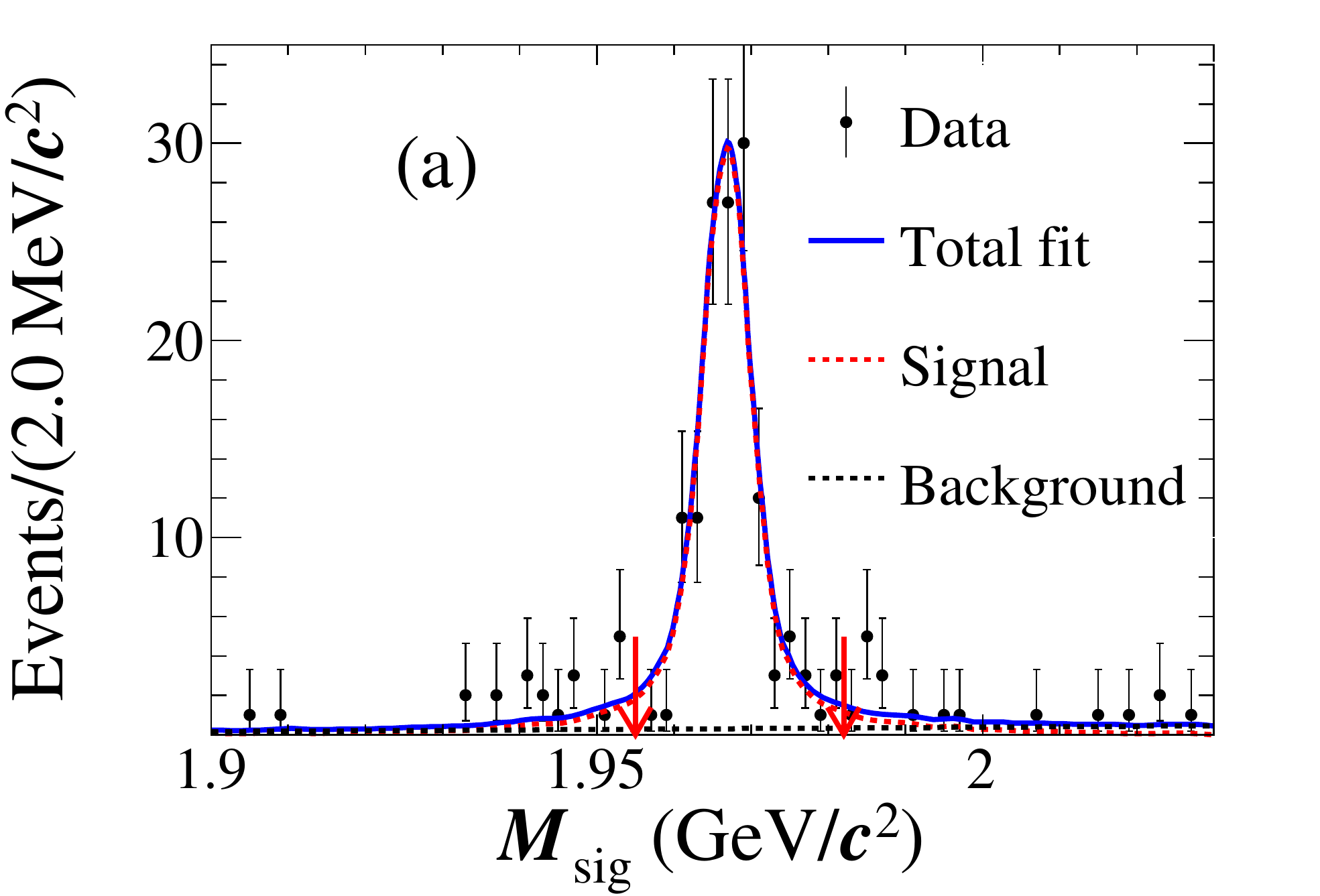}\\
  \includegraphics[width=6.cm]{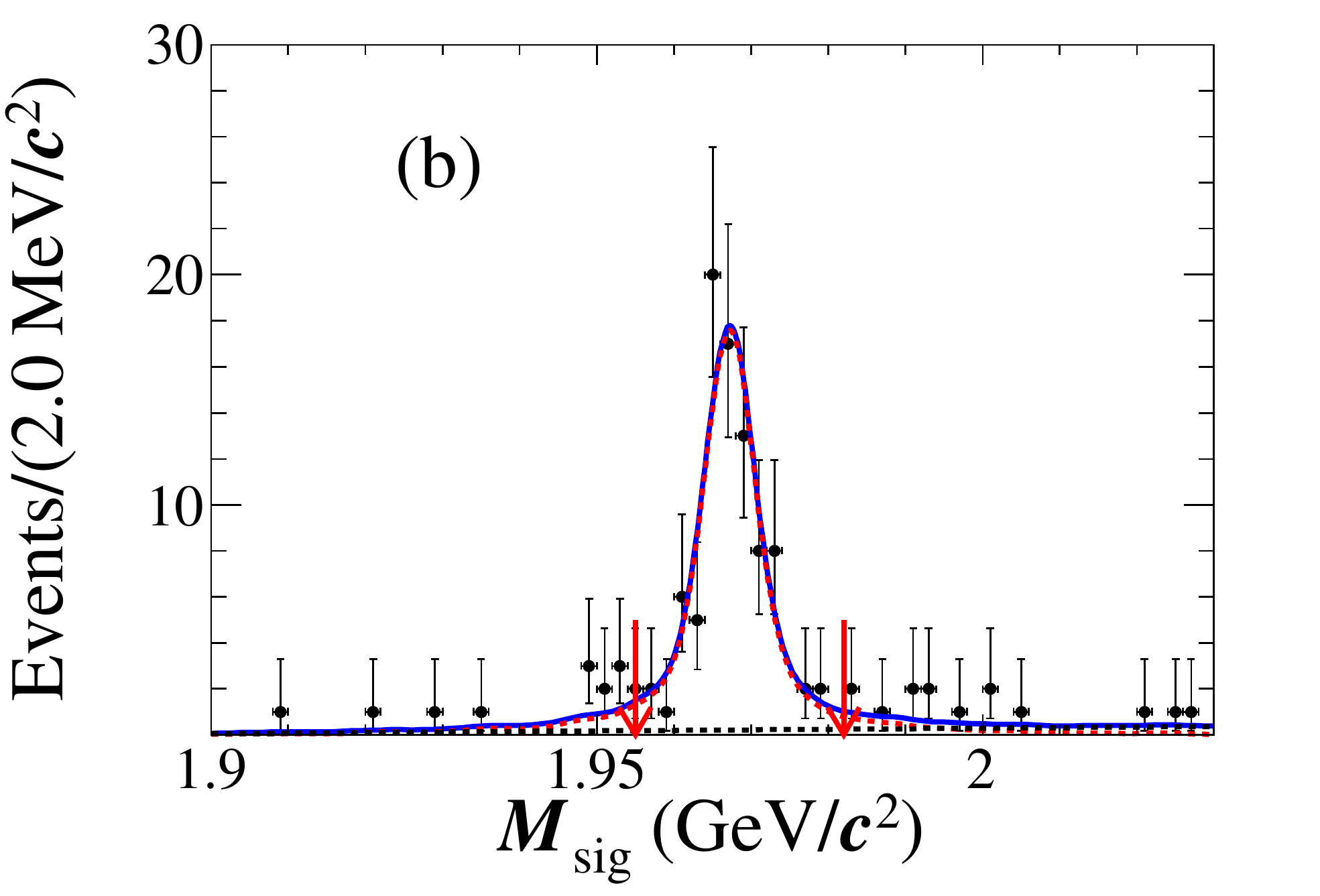}
  \includegraphics[width=6.cm]{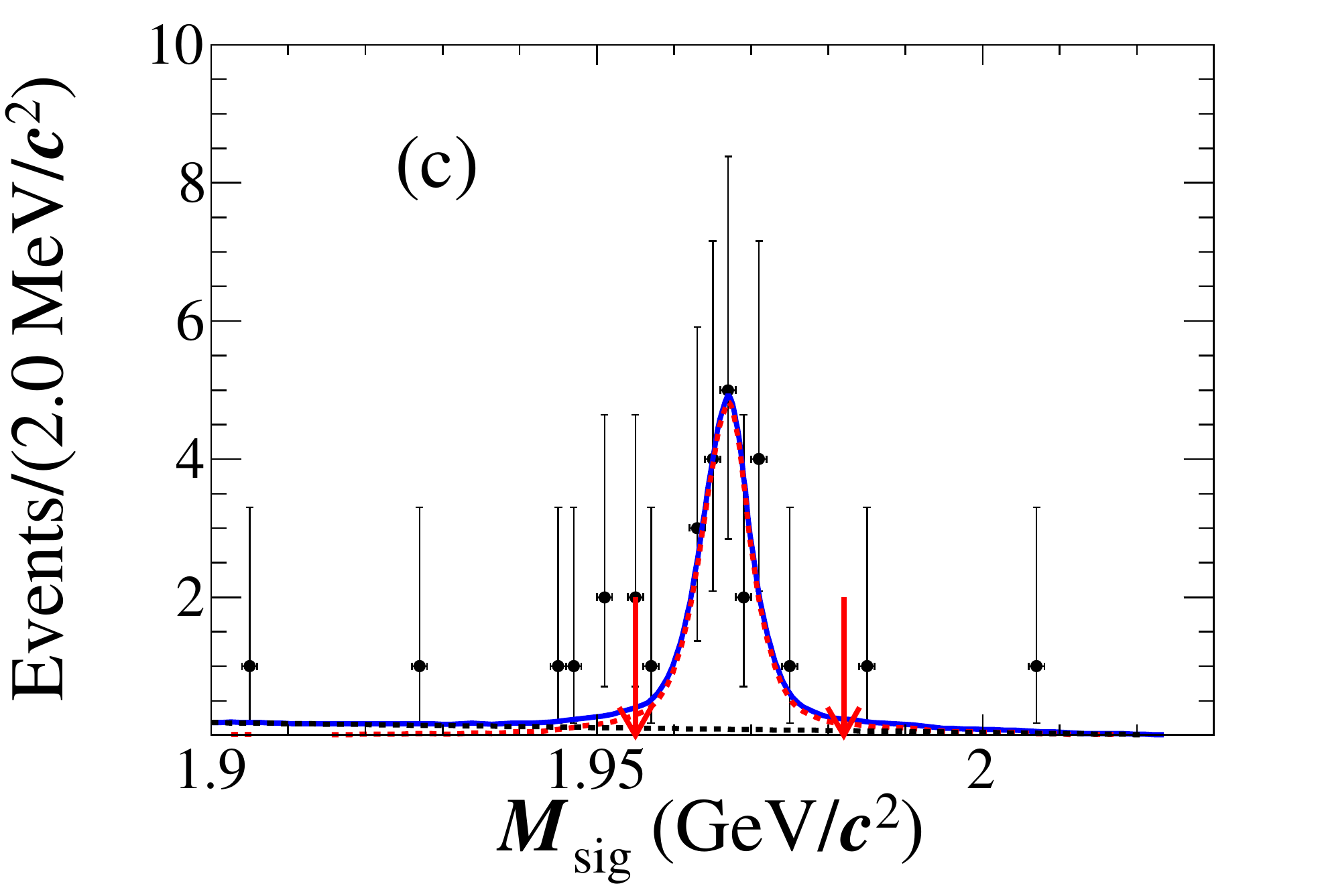}
  \caption{
    Fits to the $M_{\rm sig}$ distributions of the data samples at $\sqrt{s}=$ (a) 4.178~GeV, (b) 4.189-4.219~GeV and (c) 4.226~GeV. 
		The black points with error bars are data. 
		The blue solid lines are the total fits. 
		The red dotted and the black dashed lines are the fitted signal and background, respectively.    The pairs of red arrows indicate the signal regions.
  } \label{fig:fit_Ds}
\end{figure*}

Figure~\ref{fig:fit_Ds} shows the fits to the invariant-mass distributions of the accepted signal $D_s^{+}$ candidates ($M_{\rm sig}$) for various data samples. 
The signal is described by a MC-simulated shape convolved with a Gaussian function, and the background is described by a linear function.
Finally, a mass window $[1.955, 1.982]$~GeV/$c^2$ is applied on the signal $D_s^{+}$ candidates. 
There are 137, 84 and 22 events retained for the amplitude analysis with signal purities $(96.9\pm1.5)\%$, $(96.7\pm2.0)\%$ and $(94.9\pm4.7)\%$ for the data samples at $\sqrt{s}=4.178$, 4.189-4.219 and 4.226~GeV, respectively.

\subsection{Fit method}
The amplitude analysis of $D_s^{+}\to K^-K^+\pi^+\pi^+\pi^-$ decay is performed by using an unbinned maximum likelihood fit.
The isobar formulism is used to model the total amplitude. 
For the decay of $D_s^{+}\to K^-K^+\pi^+\pi^+\pi^-$, there are three intermediate resonances at most. 
For example, $D_s \to R_1 R_2, R_1 \to R_3 P_1, R_3 \to P_2 P_3, R_2 \to P_4 P_5$, where $R_1, R_2$, and $R_3$ are intermediate resonances.
The amplitude of the $n^{\rm th}$ intermediate process ($\mathcal A_n$) is given by:
\begin{equation}
  \mathcal A_n(p_j) = P^1_n(m_1)P^2_n(m_2)P^3_n(m_3)S_n(p_j)F^1_n(p_j)F^2_n(p_j)F^3_n(p_j)F^{D_s}_n(p_j),
\end{equation}
where $p_j$ is the set of the final state particles' four momenta, 
the index $j$ refers to the different particles in the final states, 
the indices 1, 2 and 3 correspond to the three intermediate resonances. 
In the amplitude, $S_n(p_j)$ is the spin factor, 
$F^{1,2,3}_n(p_j)$ and $F^{D_s}_n(p_j)$ are the Blatt-Weisskopf barrier factors for the intermediate resonances and $D_s^+$, 
and $P^{1,2,3}$ are the propagators for the intermediate resonances.
For the non-resonance (NR) amplitude $D_s^{+}\to (K^-K^+\pi^+\pi^+\pi^-)_{\rm NR}$, we use $\mathcal A_{\rm n}(p_j)$ = 1.
The total amplitude $\mathcal M$ is the coherent sum of the amplitudes of the intermediate processes, $\mathcal M(p_j) = \sum{c_n\mathcal A_n(p_j)}$,
where $c_n = \rho_ne^{i\phi_n}$ is the corresponding complex coefficient. 
The magnitude $\rho_n$ and phase $\phi_n$ will be determined in the likelihood fit.
The signal probability density function (PDF) $\it{f_S(p_j)}$ is given by:
\begin{equation}
\it{f_S}(p_j) = \frac{\epsilon(p_j)|\mathcal M(p_j)|^2R_5(p_j)}{\int \epsilon(p_j)|\mathcal M(p_j)|^2R_5(p_j)dp_j},
\label{pwa:pdf}
\end{equation}
where $\epsilon(p_j)$ is the detection efficiency parameterised in terms of the final four-momenta $p_j$ and $R_5(p_j)$ is the PHSP element of five-body decays.  
In the numerator of Eq.~(\ref{pwa:pdf}), $\epsilon(p_j)$ and $R_5(p_j)$ terms are independent of the fitted variables, so they are regarded as constant terms in the fit.
The normalization integrals are determined by a MC integration:
\begin{equation}
\int \epsilon(p_j)|\mathcal M(p_j)|^2R_5(p_j)\,dp_{j} \approx \frac{1}{N_{\rm gen}}\sum^{N_{\rm MC}}_{k_{\rm MC}}\frac{|\mathcal M(p^{k_{\rm MC}}_j)|^2}{|\mathcal M^{\rm gen}(p^{k_{\rm MC}}_j)|^2},
\end{equation}
where $k_{\rm MC}$ is the index of the $k^{\rm th}_{\rm MC}$ event of the MC sample, $N_{\rm gen}$ is the number of the generated MC events and $N_{\rm MC}$ is the number of the selected MC events. 
The $\mathcal M^{\rm gen}(p_j)$ is the PDF used to generate the MC samples in the MC integration.
The computational efficiency of the MC integration is significantly improved by evaluating the normalization integral with signal MC samples, which intrinsically take into account the event selection acceptance and the detection resolution. 

The effect from the tracking and PID differences between data and simulation is considered by multiplying the weight of the MC event by a factor $\gamma_{\epsilon}$, which is calculated as:
\begin{equation}
  \gamma_{\epsilon}(p_j) = \prod_{i} \frac{\epsilon_{i,\rm data}(p_j)}{\epsilon_{i,\rm MC}(p_j)},
  \label{pwa:gamma}
\end{equation}
where $i$ refers to tracking or PID, $\epsilon_{i,\rm data}(p_j)$ and $\epsilon_{i,\rm MC}(p_j)$ is the tracking or PID efficiency as a function of the momenta of the daughter particles for data and MC, respectively.
By weighting each signal MC event with $\gamma_{\epsilon}$, the MC integration is given by:
\begin{equation}
  \int \epsilon(p_j)|\mathcal M(p_j)|^2R_5(p_j)dp_j \approx \frac{1}{N_{\rm MC}}\sum^{N_{\rm MC}}_{k_{\rm MC}}\frac{\gamma_{\epsilon}(p_j^{k_{\rm MC}})|\mathcal M(p^{k_{\rm MC}}_j)|^2}{|\mathcal M^{\rm gen}(p^{k_{\rm MC}}_j)|^2}.\label{likelihood3}
\end{equation}

The contribution from the background is subtracted in the likelihood calculation by assigning a negative weight to the background events.
The log-likelihood function is written as:
\begin{equation}
  \ln{\mathcal{L}} = \begin{matrix}\sum\limits_{k}^{N_{\rm data}} \ln f_{S}(p_{j}^{k})-\sum\limits_{k'}^{N_{\rm bkg}} w_{k'}^{\rm bkg}\ln f_{S}(p_{j}^{k'})\end{matrix},  \label{loglikelihood}
\end{equation}
where $N_{\rm data}$ is the number of candidate events in data, $w_{k'}^{\rm bkg}$ and $N_{\rm bkg}$ are the background weight and the number of simulated background events, respectively.

 To combine the data samples taken at various center-of-mass energies, Eq.~(\ref{loglikelihood}) is re-written as:
\begin{equation}
  \ln{\mathcal{L}} = \begin{matrix}\sum\limits_{n=1}^{3} \ln{\mathcal{L}_n}\end{matrix},  \label{likelihood2}
\end{equation}
where n denotes the data samples at $\sqrt{s} = 4.178$ GeV, 4.189-4.219 GeV, and 4.226 GeV, respectively.

\subsubsection{Blatt-Weisskopf barriers}
\label{Blatt-Weisskopfbarriers}
For a decay process $a \to bc$, the Blatt-Weisskopf barriers $X_{L}(q)$~\cite{BESIII:2020ctr} depend on the angular momentum $L = 0, 1, 2$ and the momentum $q$ of the final-state particle $b$ or $c$ in the rest system of $a$. 
They are defined as:
\begin{equation}
\begin{aligned}
  &X_{L=0}(q)=1,\\
  &X_{L=1}(q)=\sqrt{\frac{z_0^2+1}{z^2+1}},\\
  &X_{L=2}(q)=\sqrt{\frac{z_0^4+3z_0^2+9}{z^4+3z^2+9}}, \label{xl}
\end{aligned}
\end{equation}
with $z_0 = q_0R$ and $z = qR$, where $R$ is the effective radius of the intermediate resonances. 
The momentum $q$ is given by:
\begin{equation}
	q = \sqrt{\frac{(s_a+s_b-s_c)^2}{4s_a}-s_b}, \label{q2}
\end{equation}
where $s_a, s_b$ and $s_c$ refer to the squared invariant masses of particles $a, b$ and $c$, respectively. 
The value of $q_0$ is that of $q$ when $s_a = m_a^2$.
The effective radius of barrier $R$ is fixed to be 3.0 GeV$^{-1}$ for the intermediate resonances and 5.0 GeV$^{-1}$ for the $D_s^+$ meson. 

\subsubsection{Propagator}
\label{Propagator}
The intermediate resonances $a_1(1260)$ and $\phi$ are parameterised with the relativistic Breit-Wigner (RBW) formula:
\begin{eqnarray}
\begin{aligned}
	\begin{array}{lr}
		&P(m) = \frac{1}{(m^2_0-m^2)-im_0\Gamma(m)}, \\ 
	\end{array}
\end{aligned}
\end{eqnarray}
where $m=\sqrt{E^2-p^2}$, $m_0$ is the nominal mass of the intermediate resonance, and $\Gamma(m)$ is given by:
\begin{eqnarray}
	\begin{aligned}
  \begin{array}{lr}
		&\Gamma(m)=\Gamma_0\left(\frac{q}{q_0}\right)^{2L+1}\left(\frac{m_0}{m}\right)\left(\frac{X_L(q)}{X_L(q_0)}\right)^2,  \label{propagator}
	\end{array}
		\end{aligned}
\end{eqnarray}
where $\Gamma_0$ is the width of the intermediate resonance. 

The $\rho^0$ meson is parameterised with the Gounaris-Sakurai (GS) line shape~\cite{Gounaris:1968mw}, which is given by:
\begin{equation}
P_{\rm GS}(m)=\frac{1+d\frac{\Gamma_0}{m_0}}{(m^2_0-m^2)+f(m)-im_0\Gamma(m)},
\end{equation}
where:
\begin{equation}
	f(m)=\Gamma_0\frac{m^2_0}{q^3_0}\left[q^2(h(m)-h(m_0))+(m^2_0-m^2)q^2_0\frac{dh}{d(m^2)}\Big|_{m^2=m^2_0}\right],
\end{equation}
and the function $h(m)$ is defined as:
\begin{equation}
h(m)=\frac{2}{\pi}\frac{q}{m}\ln\left(\frac{m+2q}{2m_{\pi}}\right),
\end{equation}
with:
\begin{equation}
\frac{dh}{d(m^2)}\Big|_{m^2=m_0^2} = h(m_0)[(8q^2_0)^{-1}-(2m^2_0)^{-1}]+(2\pi m^2_0)^{-1}.
\end{equation}
The normalization condition at $P_{\rm GS}(0)$ fixes the parameter $d=\frac{f(0)}{\Gamma_0m_0}$, which results in:
\begin{equation}
d=\frac{3}{\pi}\frac{m^2_{\pi}}{q^2_0}ln\left(\frac{m_0+2q_0}{2m_{\pi}}\right)+\frac{m_0}{2\pi q_0}-\frac{m^2_{\pi}m_0}{\pi q^3_0}.
\end{equation}

\subsubsection{Spin factors}
\label{Spinfactors}
Due to the limited phase space available in the decay, we only consider the states with angular momenta up to 2.
As discussed in~\cite{covariant-tensors}, we define the spin projection operator for a process $a \to bc$, $P^{(S)}_{\mu_1\cdots\mu_S\nu_1\cdots\nu_S}$ as:
\begin{equation}
  P^{(1)}_{\mu\nu}=-g_{\mu\nu}+\frac{p_{a\mu}p_{a\nu}}{p^2_a},
\end{equation}
\begin{equation}
  P^{(2)}_{\mu_1\mu_2\nu_1\nu_2}=\frac{1}{2}(P^{(1)}_{\mu_1\nu_1}P^{(1)}_{\mu_2\nu_2}+P^{(1)}_{\mu_1\nu_2}P^{(1)}_{\mu_2\nu_1})-\frac{1}{3}P^{(1)}_{\mu_1\mu_2}P^{(1)}_{\nu_1\nu_2}.
\end{equation}

The quantities $p_a$, $p_b$, and $p_c$ are the momenta of particles $a$,
$b$, and $c$, respectively.
The covariant tensors are given by:
\begin{eqnarray}
\begin{aligned}
    \tilde{t}^{(1)}_{\mu}(a) &= -P^{(1)}_{\mu\mu^{\prime}}(a)r^{\mu^{\prime}}_{a}\,,\\
    \tilde{t}^{(2)}_{\mu\nu}(a) &= P^{(2)}_{\mu\nu\mu^{\prime}\nu^{\prime}}(a)r^{\mu{\prime}}_{a}r^{\nu^{\prime}}_{a}\,.\\
\label{covariant-tensors}
\end{aligned}
\end{eqnarray}
where $r_a = p_b-p_c$.

Eleven kinds of spin factors are listed in Table~\ref{pwa:spin}, where the tensor describing the $D_{s}^{+}$ decay is denoted by $\tilde{T}$ and the one of the $a$ decay is denoted by $\tilde{t}$. 

\begin{table}[htbp]
	\caption{Spin factor for each decay chain. $[S]$, $[P]$, and $[D]$ indicate the orbital angular momenta $L$ = 0, 1, and 2 of the two-body final states, respectively.}
  \centering
	\footnotesize
  \begin{tabular}{l|c}
  \hline
  \hline
    Decay chain   &Spin factor \\
  \hline
    $D_s[S]\to AV_{1},A[S]\to V_{2}P_1,V_{1}\to P_2P_3,V_{2}\to P_4P_5$ &$P_{(1)}^{\mu\nu}(A)\tilde{t}_{(1)\mu}(V_{1})\tilde{t}_{(1){\nu}}(V_{2})$\\ \hline
    $D_s[S]\to AV_{1},A[D]\to V_{2}P_1,V_{1}\to P_2P_3,V_{2}\to P_4P_5$ &$\tilde{t}_{(2)}^{\mu\nu}(A)\tilde{t}_{(1)\mu}(V_{1})\tilde{t}_{(1){\nu}}(V_{2})$\\ \hline
    $D_s[P]\to AV_{1},A[S]\to V_{2}P_1,V_{1}\to P_2P_3,V_{2}\to P_4P_5$ &$\epsilon_{\mu\nu\lambda\sigma}p^{\mu}(D)\tilde{T}^{(1)\nu}(D)P_{(1)}^{\beta\lambda}(A)\tilde{t}_{(1){\beta}}(V_{2})\tilde{t}^{(1)\sigma}(V_{1})$\\ \hline
    $D_s[P]\to AV_{1},A[D]\to V_{2}P_1,V_{1}\to P_2P_3,V_{2}\to P_4P_5$ &$\epsilon_{\mu\nu\lambda\sigma}p^{\mu}(D)\tilde{T}^{(1)\nu}(D)\tilde{t}_{(2)}^{\beta\lambda}(A)\tilde{t}_{(1){\beta}}(V_{2})\tilde{t}^{(1)\sigma}(V_{1})$\\ \hline
    $D_s[D]\to AV_{1},A[S]\to V_{2}P_1,V_{1}\to P_2P_3,V_{2}\to P_4P_5$ &$\tilde{T}_{(2)\mu\nu}(D)P_{(1)}^{\mu\beta}(A)\tilde{t}_{(1)\beta}(V_{2})\tilde{t}^{(1){\nu}}(V_{1})$\\ \hline
    $D_s[D]\to AV_{1},A[D]\to V_{2}P_1,V_{1}\to P_2P_3,V_{2}\to P_4P_5$ &$\tilde{T}_{(2)\mu\nu}(D)\tilde{t}_{(2)}^{\mu\beta}(A)\tilde{t}_{(1)\beta}(V_{2})\tilde{t}^{(1){\nu}}(V_{1})$\\ \hline
    $D_s[S]\to AV,A[P]\to SP_1,V\to P_2P_3,S\to P_4P_5$ &$\tilde{t}_{(1)}^{\mu}(A)\tilde{t}_{(1)\mu}(V)$\\ \hline
    $D_s[P]\to AV,A[P]\to SP_1,V\to P_2P_3,S\to P_4P_5$ &$\epsilon_{\mu\nu\lambda\sigma}p^{\mu}(D)\tilde{T}^{(1)\nu}(D)\tilde{t}_{(1)}^{\lambda}(A)\tilde{t}^{(1)\sigma}(V)$\\ \hline
    $D_s[D]\to AV,A[P]\to SP_1,V\to P_2P_3,S\to P_4P_5$ &$\tilde{T}_{(2)\mu\nu}(D)\tilde{t}^{(1)\mu}(A)\tilde{t}^{(1){\nu}}(V)$\\\hline
    $D_s[P]\to AS,A[S]\to VP_1,S\to P_2P_3,V\to P_4P_5$ &$\tilde{T}_{(1)\beta}(D)P_{(1)}^{\beta\nu}(A)\tilde{t}_{(1){\nu}}V$\\ \hline
    $D_s[P]\to AS,A[D]\to VP_1,S\to P_2P_3,V\to P_4P_5$ &$\tilde{T}_{(1)\beta}(D)\tilde{t}_{(2)}^{\beta\nu}(A)\tilde{t}_{(1){\nu}}V$\\
    \hline
  \hline
  \end{tabular}
  \label{pwa:spin}
\end{table}

\subsection{Fit results}
The amplitude of the $D_s^+[S]\to a_1(1260)^+\phi, a_1(1260)^+[S]\to\rho^0\pi^+, \phi\to K^-K^+$ decay is expected to have the largest contribution and it has been chosen as the reference. 
Thus, its magnitude and phase are fixed to 1.0 and 0.0, respectively, while the other amplitudes are left floating in the amplitude fit. 
The masses and widths of all the resonances are fixed to the corresponding PDG averages~\cite{PDG}. 
The background weights are fixed according to the fits shown in Fig.~\ref{fig:fit_Ds}.

We first consider the two amplitudes $D_s^+[S]\to a_1(1260)^+\phi, a_1(1260)^+[S]\to\rho^0\pi^+, \phi\to K^-K^+$ and $D_s^{+}\to (K^-K^+\pi^+\pi^+\pi^-)_{\rm NR}$. 
We have tested all the possible processes including $D_s^+\to \phi \pi^+\pi^+\pi^-$, $D_s^+\to K_1(1270)^{+}\bar K^{*0}(892)$ and $D_s^+\to a_1(1260)^+a_0(980)$, which are listed in Appendix~\ref{text},
and only the amplitudes of $D_s^+[S]\to a_1(1260)^+\phi, a_1(1260)^+[S]\to\rho^0\pi^+, \phi\to K^-K^+$, $D_s^+[P]\to$ $a_1(1260)^+\phi, a_1(1260)^+[S]\to\rho^0\pi^+, \phi\to K^-K^+$ and $D_s^{+}\to (K^-K^+\pi^+\pi^+\pi^-)_{\rm NR}$ have statistical significances larger than $5\sigma$ and are retained in the nominal solution.
The statistical significance of each new amplitude is calculated from the change of log-likelihood taking into account the change of degrees of freedom.

The calculation of the fit fraction~(FF) for each individual amplitude does not involve detector acceptance or resolution effects, and is based on PHSP MC truth information, as it follows:
\begin{eqnarray}\begin{aligned}
	{\rm FF}_{n} = \frac{\sum^{N_{\rm gen}} \left|\tilde A_{n}\right|^{2}}{\sum^{N_{\rm gen}} \left|\mathcal M\right|^{2}}\,, \label{Fit-Fraction-Definition}
\end{aligned}\end{eqnarray}
where $N_{\rm gen}$ is the number of PHSP MC generated events, 
$\tilde A_{n}$ is either the $n^{\rm th}$ amplitude ($\tilde A_{n} = c_n\mathcal A_{n}$) or the $n^{\rm th}$ component of a coherent sum of amplitudes ($\tilde A_{n} = \sum {c_{n,i}\mathcal A_{n,i}}$).

\begin{figure}[htp]
          \centering
          \includegraphics[width=5.8cm]{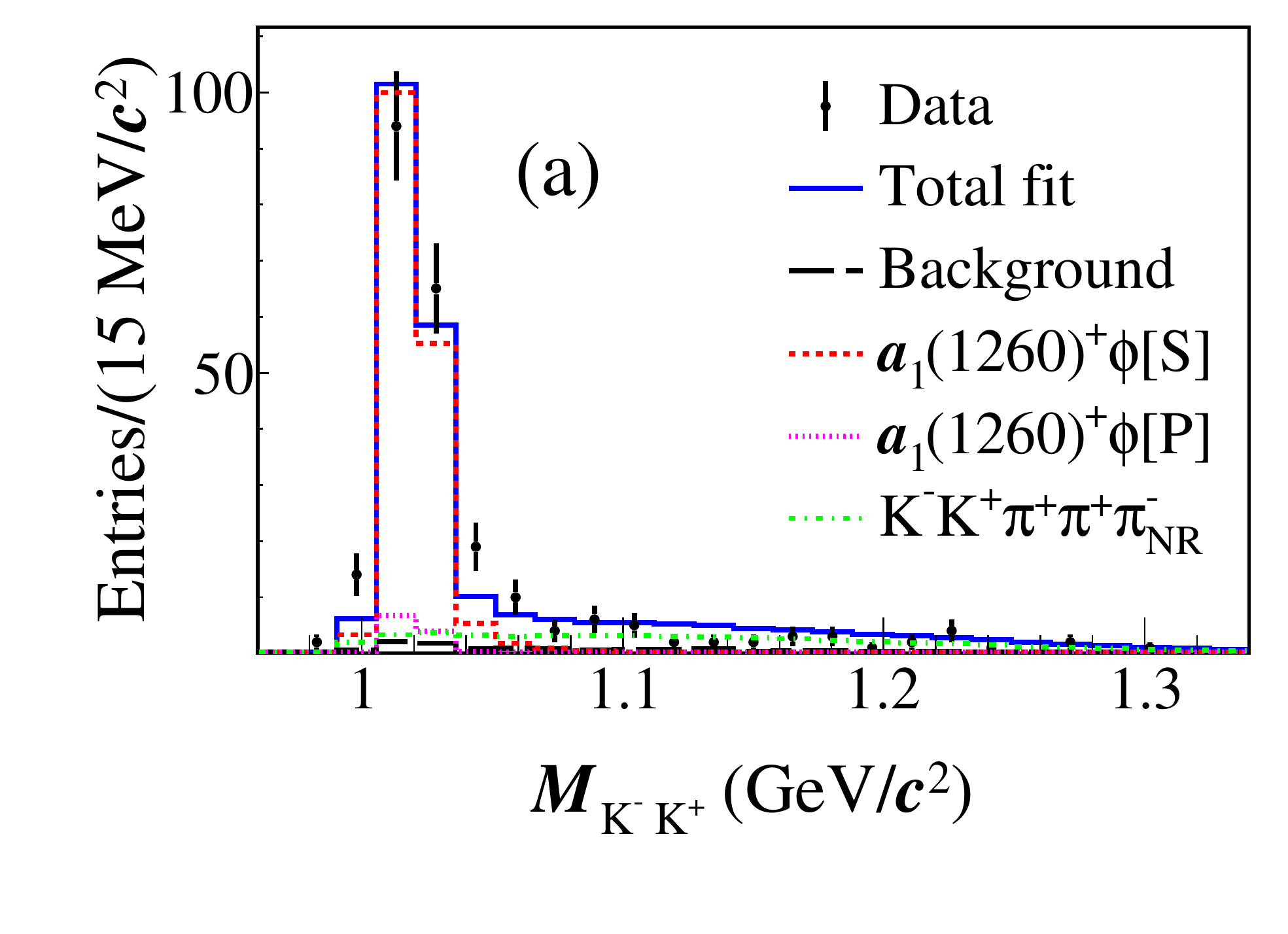}
          \includegraphics[width=5.8cm]{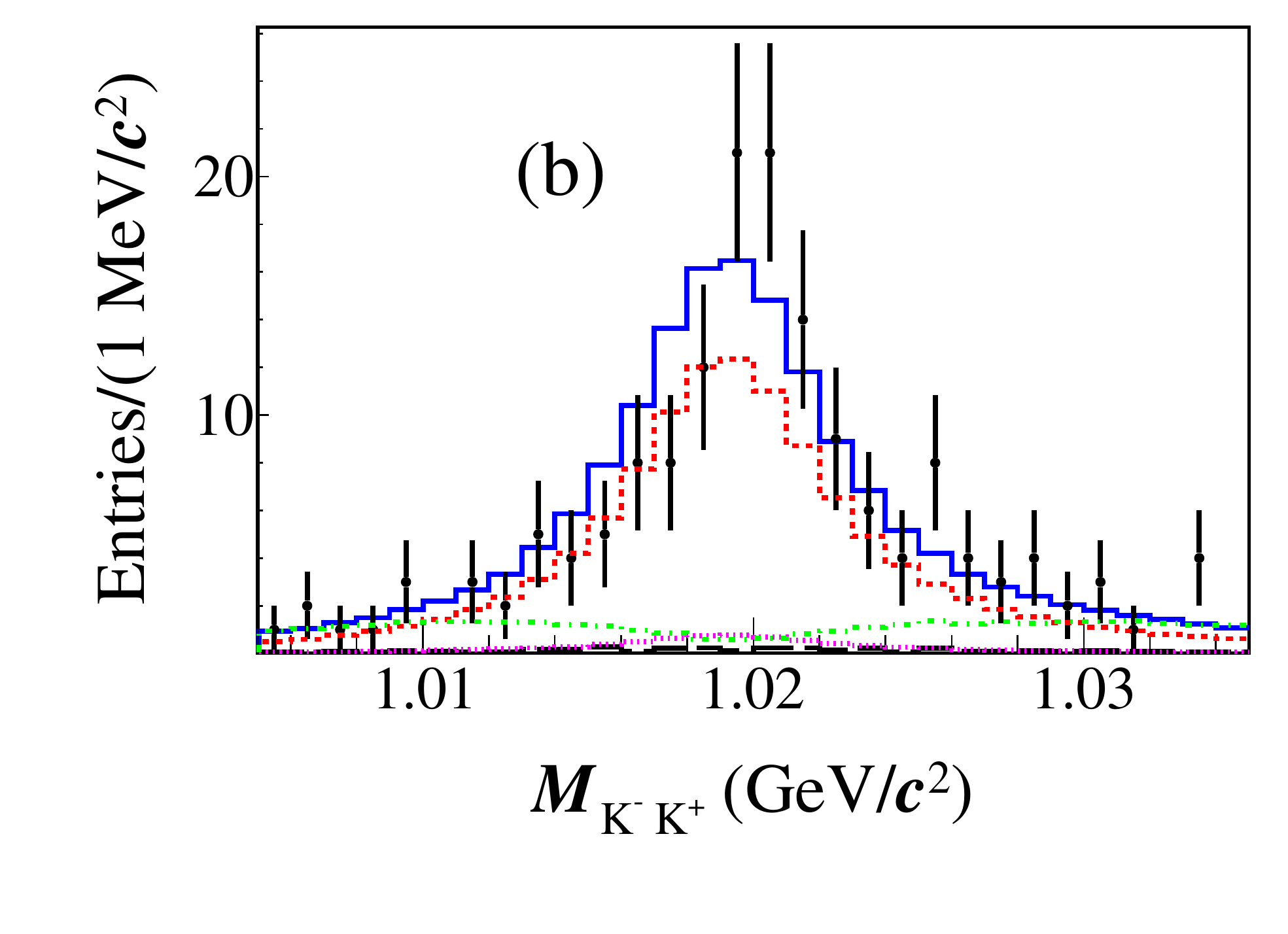}\\\
          \includegraphics[width=5.8cm]{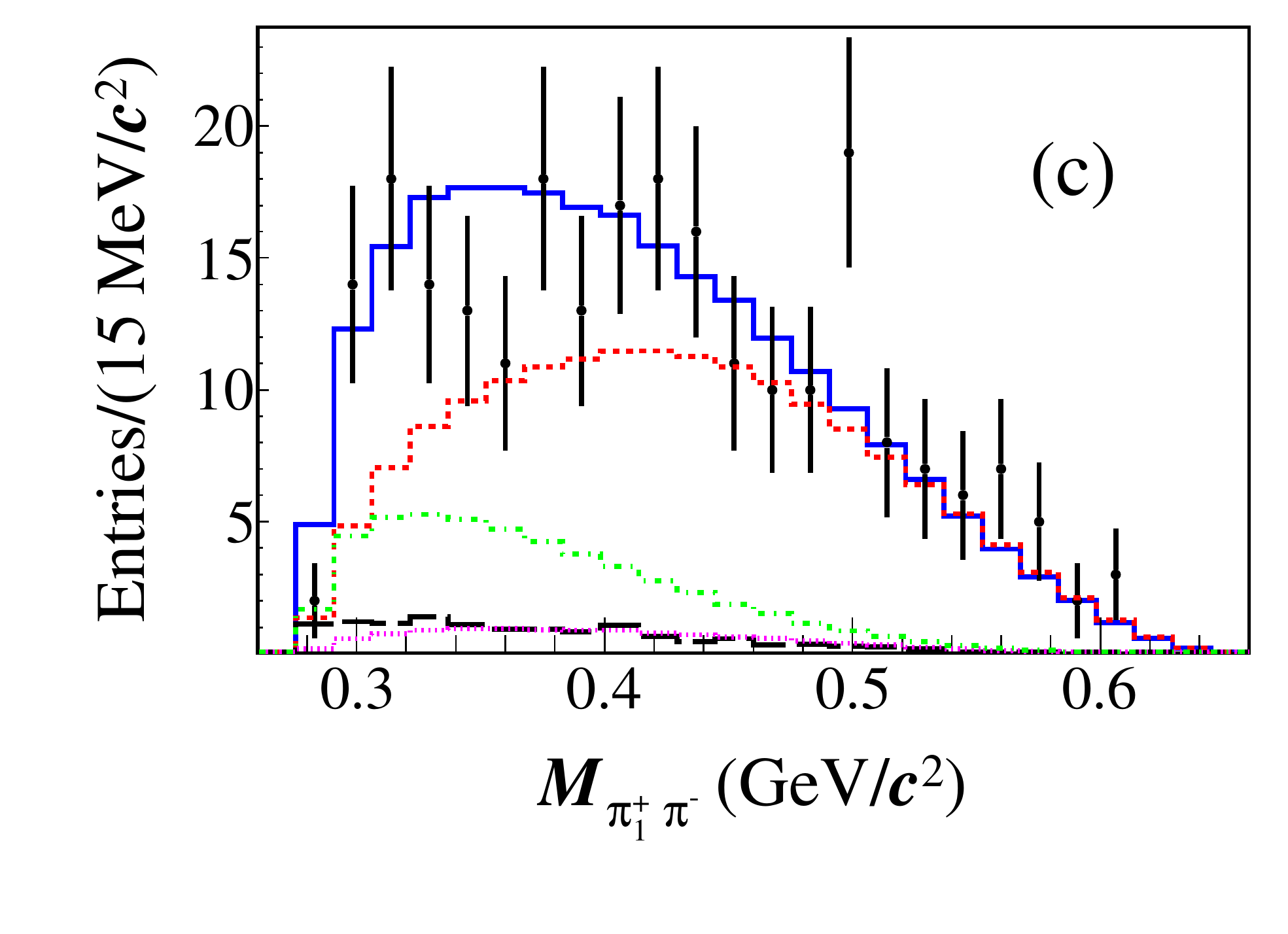}
          \includegraphics[width=5.8cm]{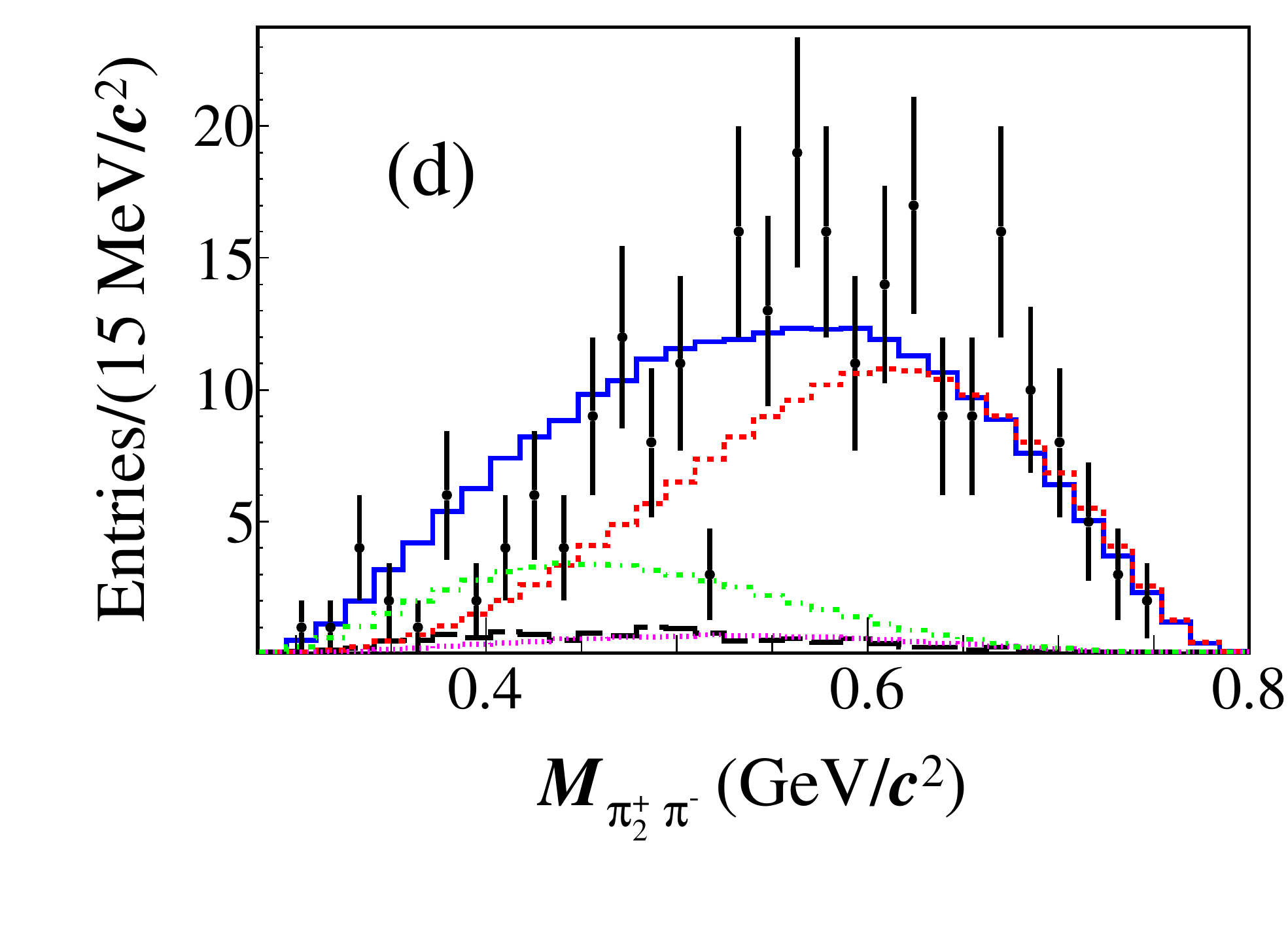}
          \includegraphics[width=5.8cm]{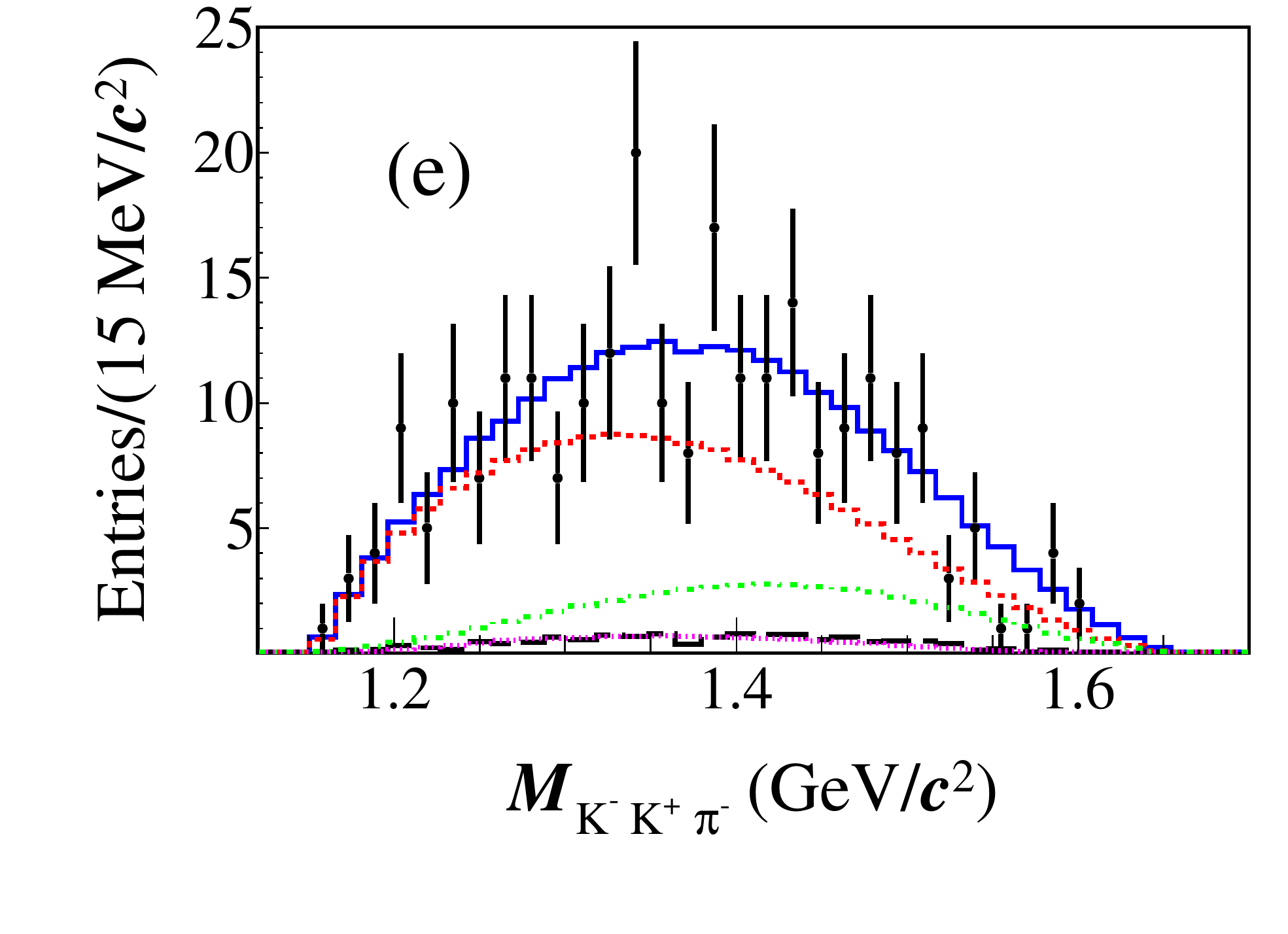}
          \includegraphics[width=5.8cm]{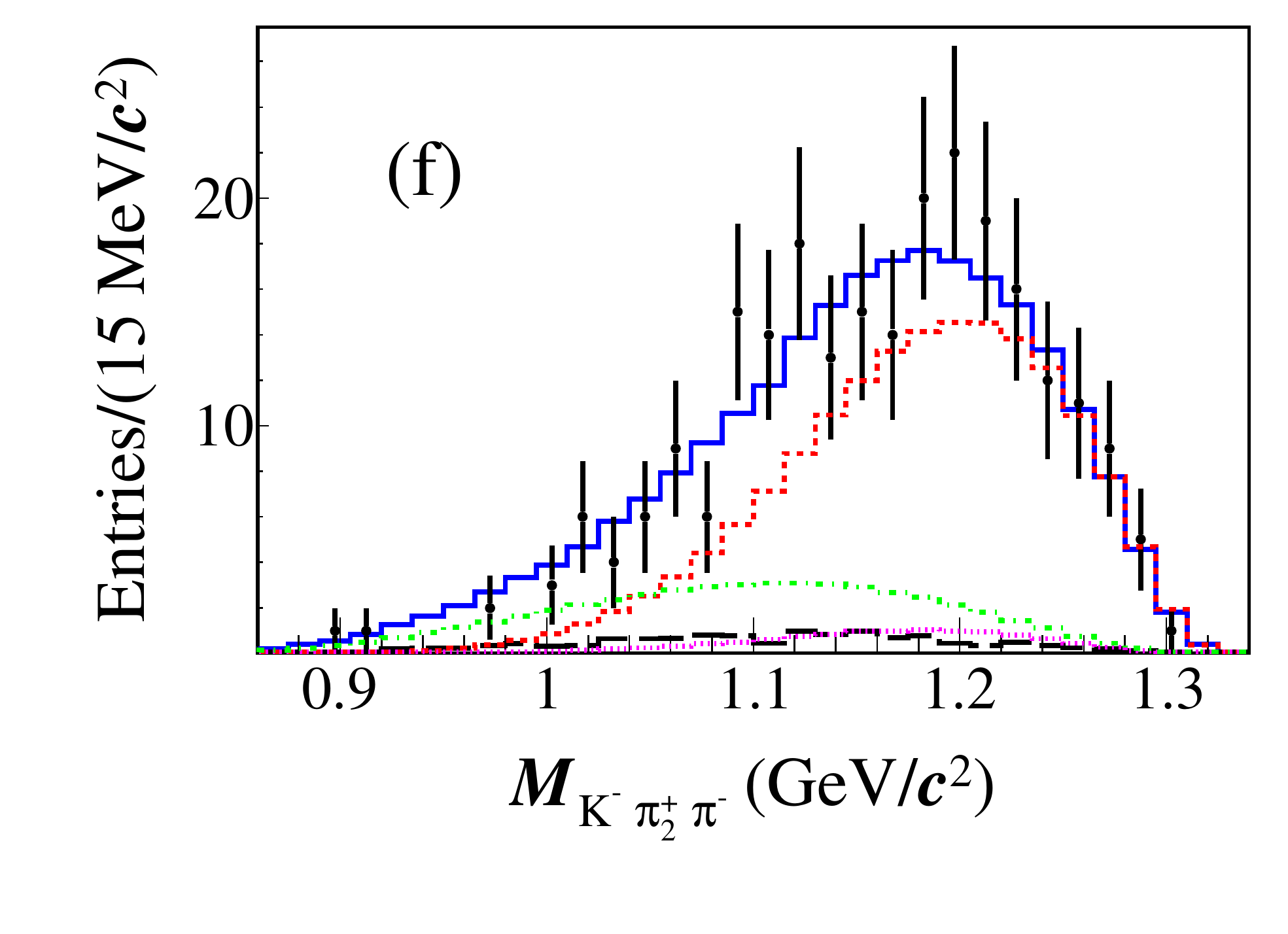}
          \includegraphics[width=5.8cm]{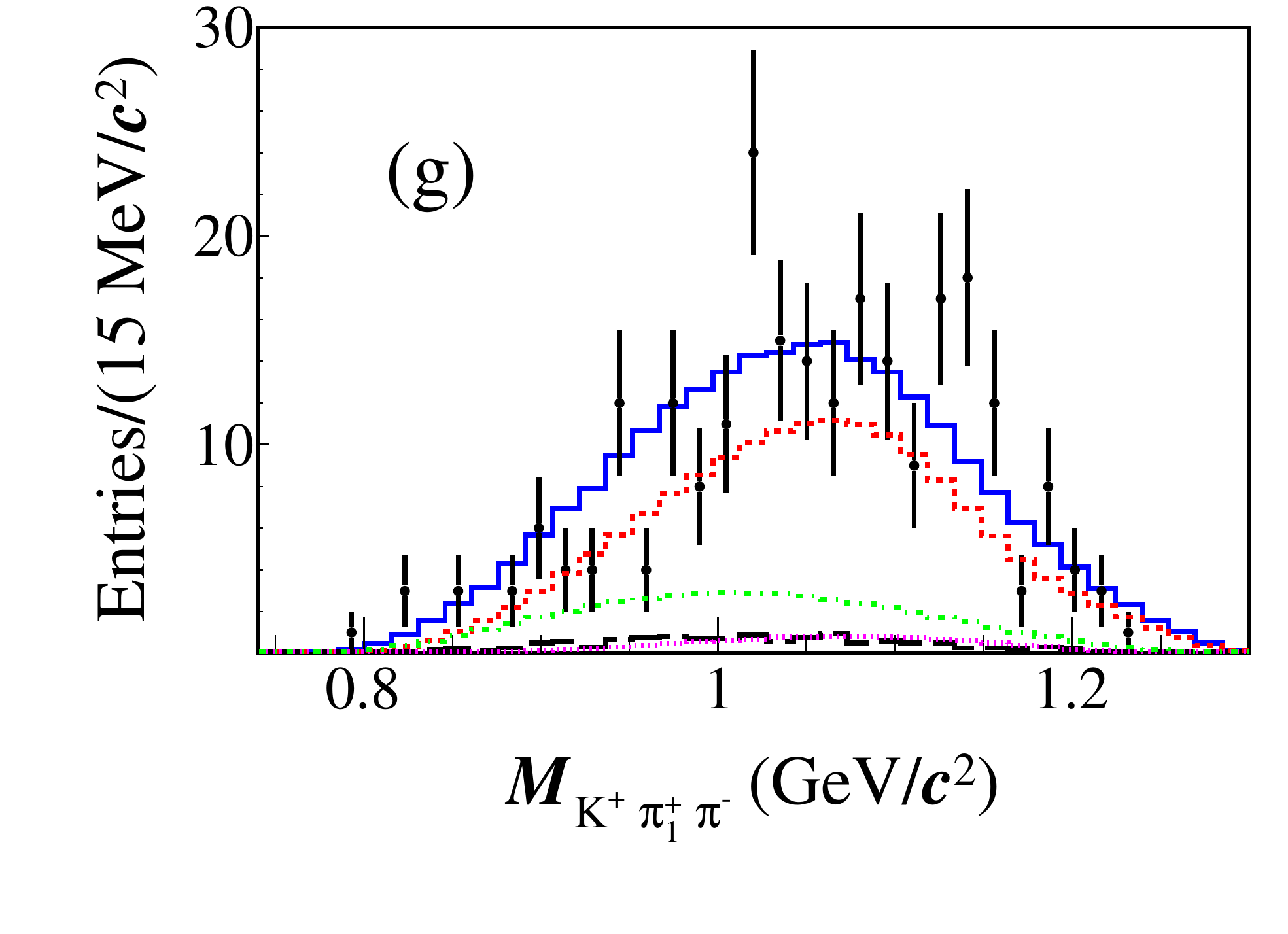}
          \includegraphics[width=5.8cm]{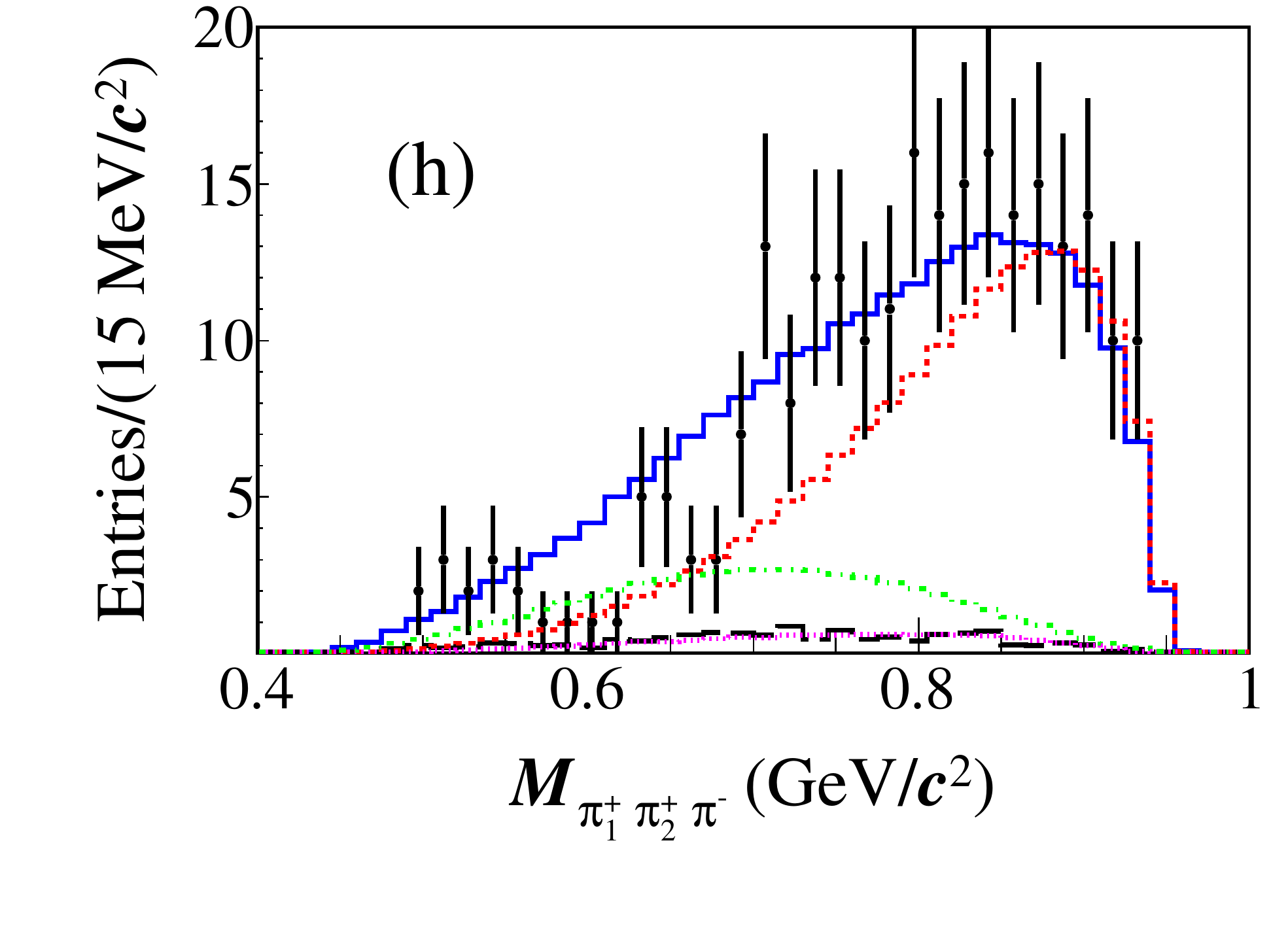}
          \includegraphics[width=5.8cm]{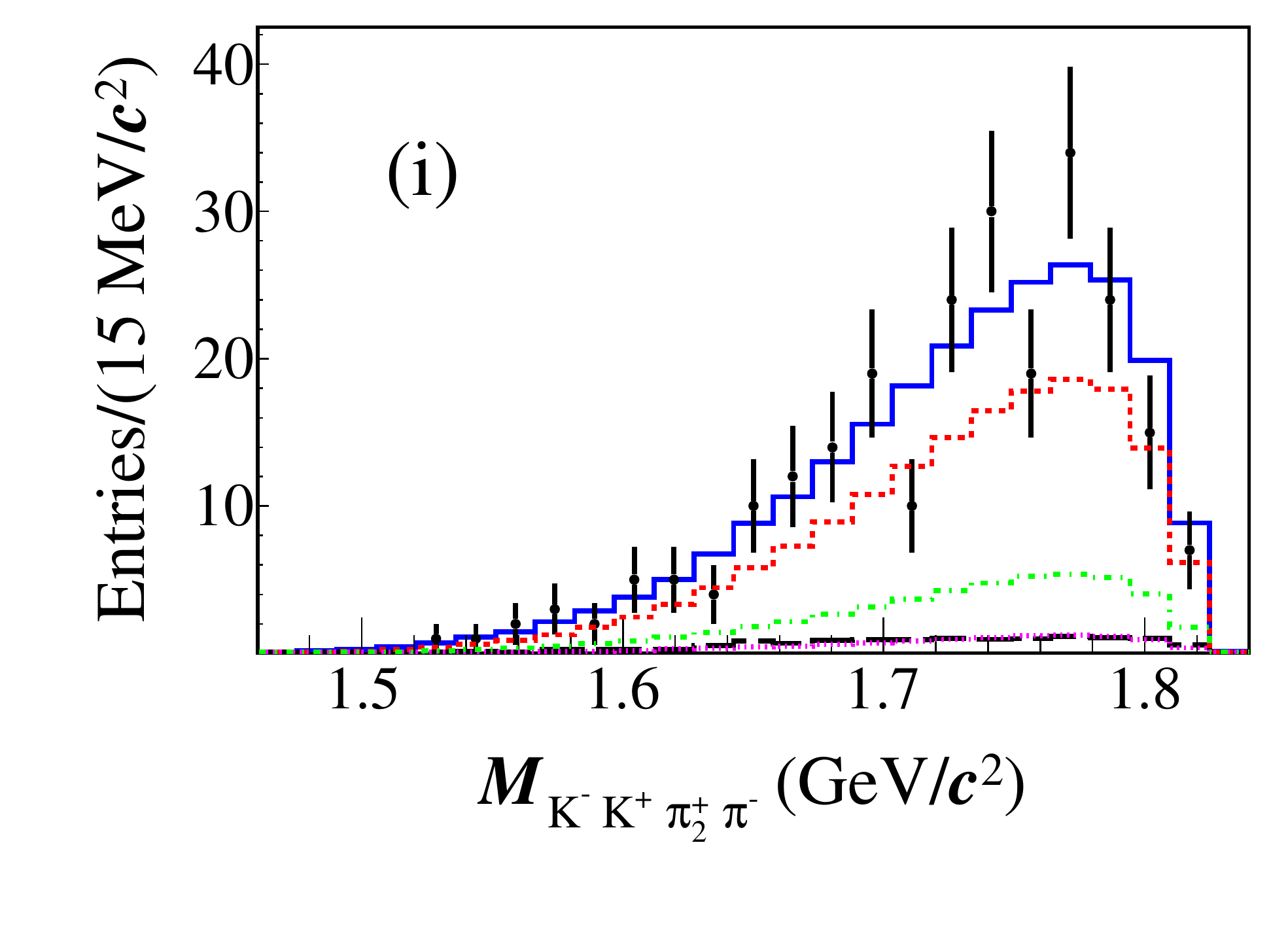}
          \includegraphics[width=5.8cm]{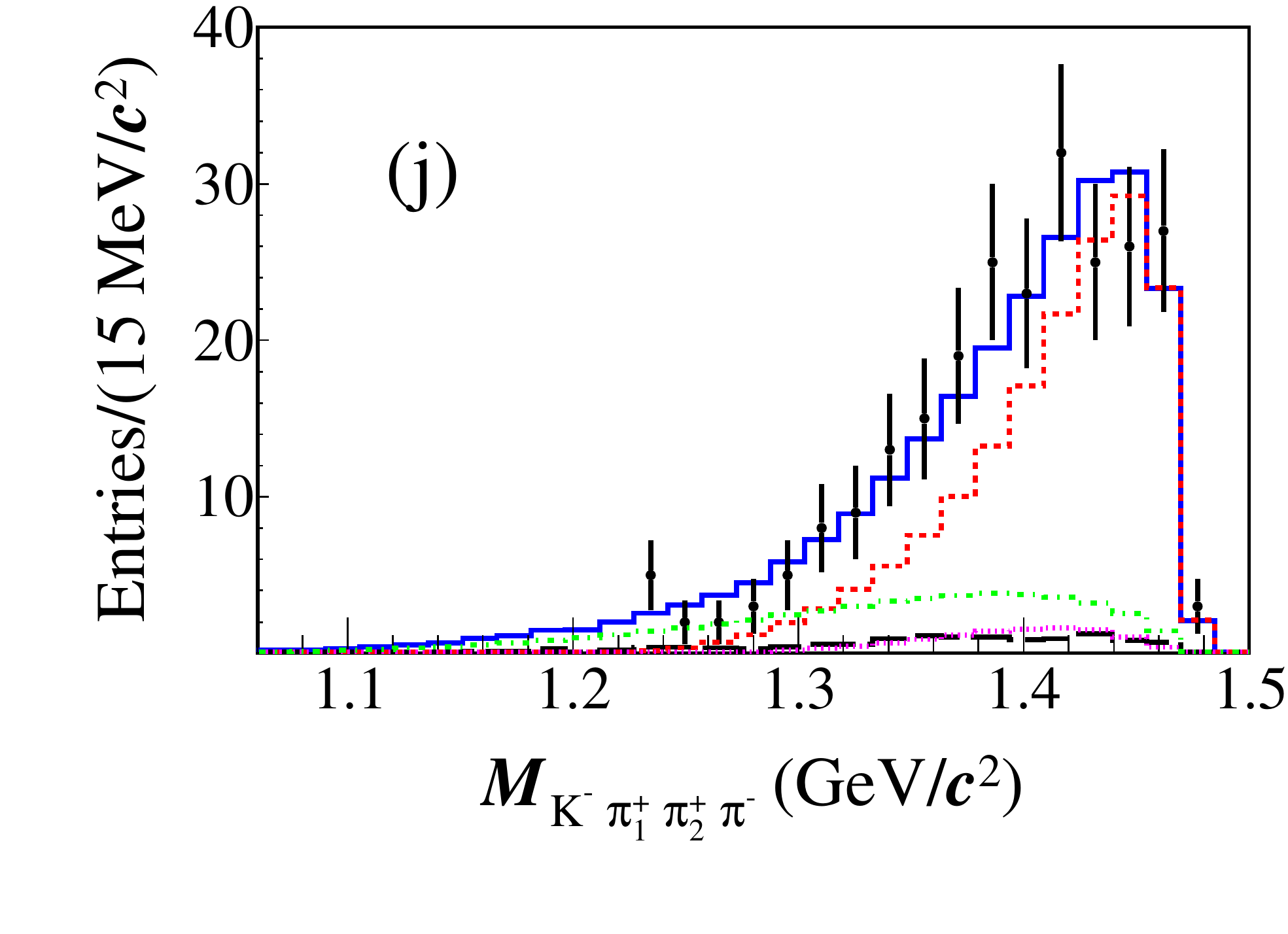}
          \caption{Mass projections from the nominal fit.
     		The data are represented by points with error bars, the fit results by the blue solid line.
     		The red dashed line is the contribution of $D_s^{+}[S]\to a_{1}(1260)^{+}\phi$,
     		the magenta dotted line is the contribution of $D_s^{+}[P]\to a_{1}(1260)^{+}\phi$,
     		the green dash-dotted line is the contribution of $D_s^{+}\to (K^-K^+\pi^+\pi^+\pi^-)_{\rm NR}$,
     		and the background estimated from the inclusive MC samples by the black long dashed line.
     		The plot (b) indicates the zoom of the $\phi$ mass region.}
    \label{pwa:proji}
\end{figure}

The phases, FFs, and statistical significances for various amplitudes are listed in Table~\ref{tab:tested_amplitudes}. 
The mass projections of fit results are shown in Fig.~\ref{pwa:proji}.
The assignments of systematic uncertainties are discussed in the next section. 

\begin{table*}[htbp]
  \caption{Phases, FFs, and statistical significances for different amplitudes. 
	The $D^+_s\to a_1(1260)^+\phi$ is the coherent sum of the $D^+_s[S]\to a_1(1260)^+\phi$ and $D^+_s[P]\to a_1(1260)^+\phi$ amplitudes. 
Due to interference effects, the total FF of amplitudes is not necessarily equal to $100\%$.
The first and the second uncertainties in the phases and FFs are statistical and systematic, respectively. 
In the table, the intermediate resonance $a_1(1260)^+[S]$ decays to $\rho^0\pi^+$.
			}
    \label{tab:tested_amplitudes}
    \begin{tabular}{ l r@{$\pm$}l r@{$\pm$}l c}
      \hline
			Amplitude                  &\multicolumn{2}{c}{Phase} &\multicolumn{2}{c}{FF (\%)} &Significance ($\sigma$) \\
			\hline
			$D^+_s[S]\to a_1(1260)^+\phi$ &\multicolumn{2}{c}{0 (fixed)}               &73.2 &$3.1\pm1.4$ &$>10$\\
      $D^+_s[P]\to a_1(1260)^+\phi$ &1.47 &$0.19\pm0.03$    &5.0   &$1.7\pm0.7$   &$5.5$\\
		  $D^+_s\to a_1(1260)^+\phi$    &\multicolumn{2}{c}{...}&78.3  &$3.5\pm1.6$   &...\\
			\hline
      $D^+_s\to (K^-K^+\pi^+\pi^+\pi^-)_{\rm NR}$                       &1.99&$0.12\pm0.05$  &21.8  &$2.9\pm0.7$   &$>10$\\
			\hline
    \end{tabular}
\end{table*}

\subsection{Systematic uncertainties for amplitude analysis}
\label{sec:PWA-Sys}
The systematic uncertainties for the amplitude analysis are summarised
in Table~\ref{systematic-uncertainties}, with their assignments described below.
\begin{itemize}
\item[\lowercase\expandafter{\romannumeral1}]
Resonance parameters: the masses and widths of $a_1(1260)$, $\phi$ and $\rho$ are shifted by their corresponding uncertainties~\cite{PDG}; 
the changes of the phases and FFs are assigned as the associated systematic uncertainties.

\item[\lowercase\expandafter{\romannumeral2}]
	$R$ values: the systematic uncertainties associated with effective radii of barriers ($R$ values) are estimated by repeating the fit procedure by varying the radii of the intermediate states and $D_s^+$ meson within 1 GeV$^{-1}$.

\item[\lowercase\expandafter{\romannumeral3}]
Fit bias: the uncertainty due to the fit procedure is evaluated by studying signal MC samples. 
An ensemble of 300 signal MC samples are generated according to the nominal results of this analysis. 
After applying the selection criteria, each of these samples has the same size as the data sample and is used to perform the same amplitude analysis. 
The pull of each parameter is defined as $\frac{\rm Out(i) - \rm In(i)}{\sigma_{\rm stat(i)}}$, where $i$ denotes the different parameters, In$(i)$ denotes the input value, Out$(i)$ is the value obtained from the fit to a signal MC sample and $\sigma_{\rm stat(i)}$ is the corresponding statistical uncertainty. 
For each parameter, 300 pull values are obtained and the deviations of their average from zero are considered as the systematic uncertainty.

\item[\lowercase\expandafter{\romannumeral4}]
  Background estimation:
  the background is determined by the inclusive MC samples; the fractions of background events are increased or decreased by one corresponding statistical uncertainties,
    and the largest differences from the nominal results are considered as the uncertainties.

\item[\lowercase\expandafter{\romannumeral5}]
	Lineshape of the $\rho$ meson: an alternative lineshape parameterization with relativistic Breit-Wigner instead of Gounaris-Sakurai is used, and differences are included inside the uncertainties.

\item[\lowercase\expandafter{\romannumeral6}]
  Experimental effects:
  to estimate the systematic uncertainty related to the difference of tracking or PID efficiencies between data and MC simulation, which is $\gamma_{\epsilon}$ in Eq.~(\ref{likelihood3}), the amplitude fit is performed varying the tracking and PID efficiencies according to their uncertainties.
  However, these uncertainties are estimated to be negligible.
  
\end{itemize}

\begin{table*}[tp]
  \renewcommand\arraystretch{1.25}
  \centering
  \caption{Systematic uncertainties on the phases and FFs for various amplitudes in units of the corresponding statistical uncertainties. The sources are:
    (\lowercase\expandafter{\romannumeral1}) the fixed parameters in the amplitudes,
    (\lowercase\expandafter{\romannumeral2}) the $R$ values,
    (\lowercase\expandafter{\romannumeral3}) fit bias,
		(\lowercase\expandafter{\romannumeral4}) background, and 
		(\lowercase\expandafter{\romannumeral5}) shape of the $\rho$ meson.
}
  \label{systematic-uncertainties}
  \begin{tabular}{lccccccc}
    \hline
    \multirow{2}{*}{Amplitude}&\multicolumn{7}{c}{Source}\cr
		& & \lowercase\expandafter{\romannumeral1} &\lowercase\expandafter{\romannumeral2} &\lowercase\expandafter{\romannumeral3} &\lowercase\expandafter{\romannumeral4} &\lowercase\expandafter{\romannumeral5} & Total   \\
    \hline
    $D^+_s[S]\to a_1(1260)^+\phi, a_{1}^{+}[S]\to \rho\pi^{+}$ &FF &0.17 &0.14 &0.18 &0.40 &0.00 &0.49\\
    \hline
    \multirow{2}{*}{$D^+_s[P]\to a_1(1260)^+\phi, a_{1}^{+}[S]\to \rho\pi^{+}$} &$\Phi$ &0.02 &0.14 &0.06 &0.01 &0.06 &0.17\\
                                                                  & FF     &0.05 &0.07 &0.12 &0.40 &0.02 &0.43\\
    \hline
    $D^+_s\to a_1(1260)^+\phi$ &FF &0.18 &0.16 &0.22 &0.57 &0.02 &0.66\\
    \hline
    \multirow{2}{*}{$D_s^{+}\to (K^-K^+\pi^+\pi^+\pi^-)_{\rm NR}$}              &$\Phi$ &0.33 &0.23 &0.08 &0.04 &0.04 &0.41\\
                                                                  & FF     &0.04 &0.10 &0.15 &0.20 &0.01 &0.27\\
    \hline
  \end{tabular}
\end{table*}

\section{Branching fraction measurement}
\label{BFSelection}
\begin{figure*}[htp]
\begin{center}
    \includegraphics[width=6.0cm]{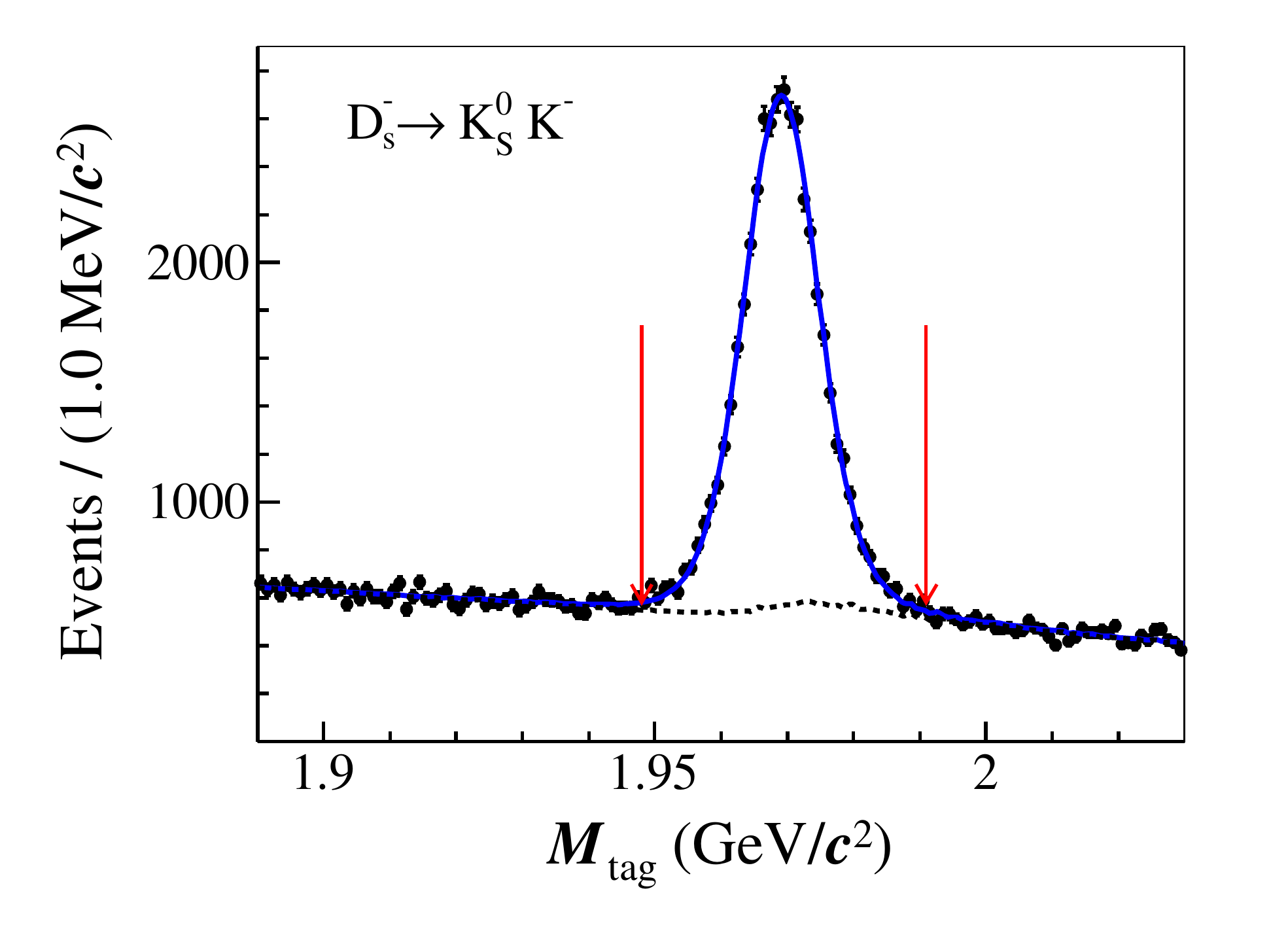}
    \includegraphics[width=6.0cm]{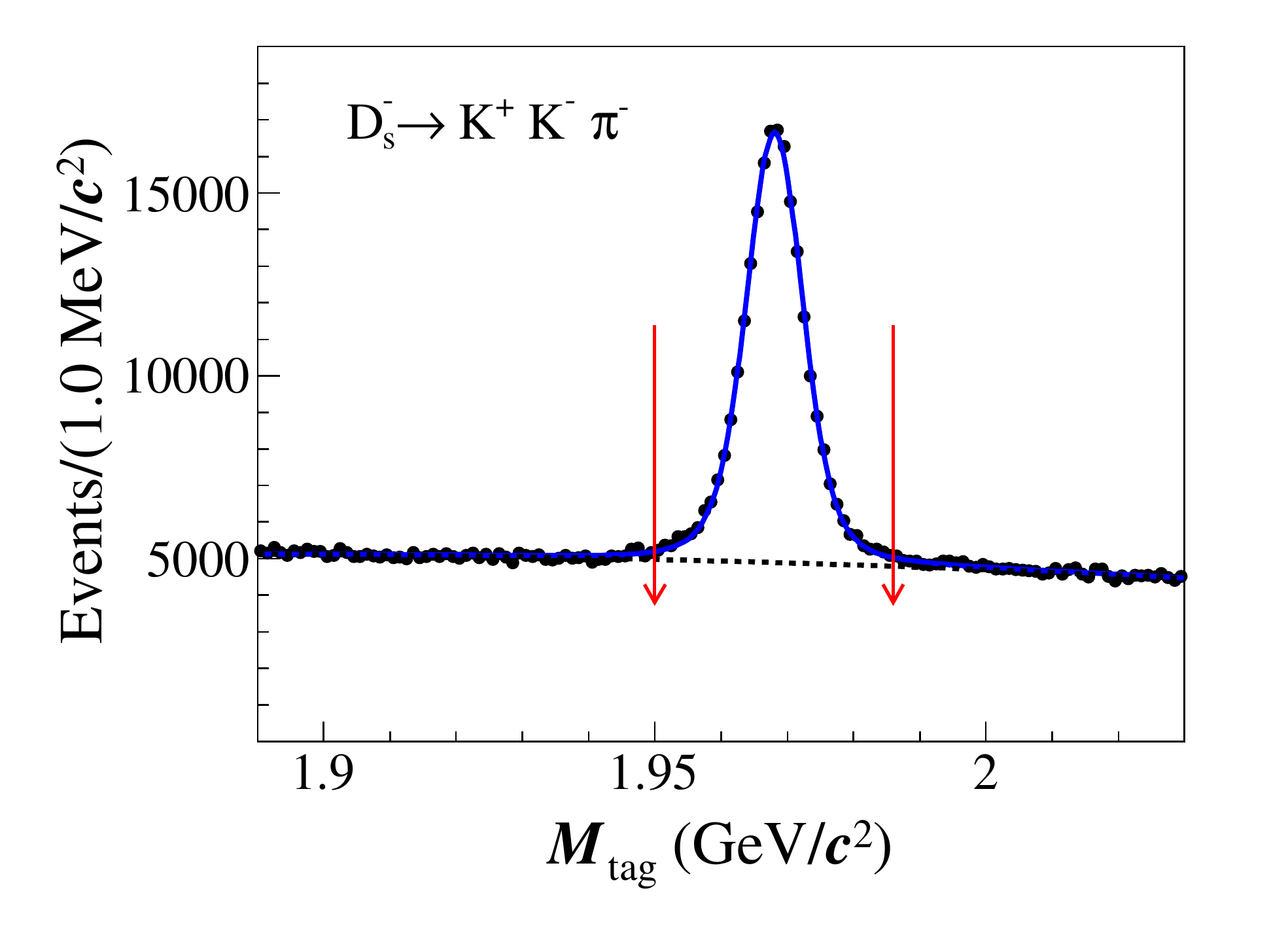}
    \includegraphics[width=6.0cm]{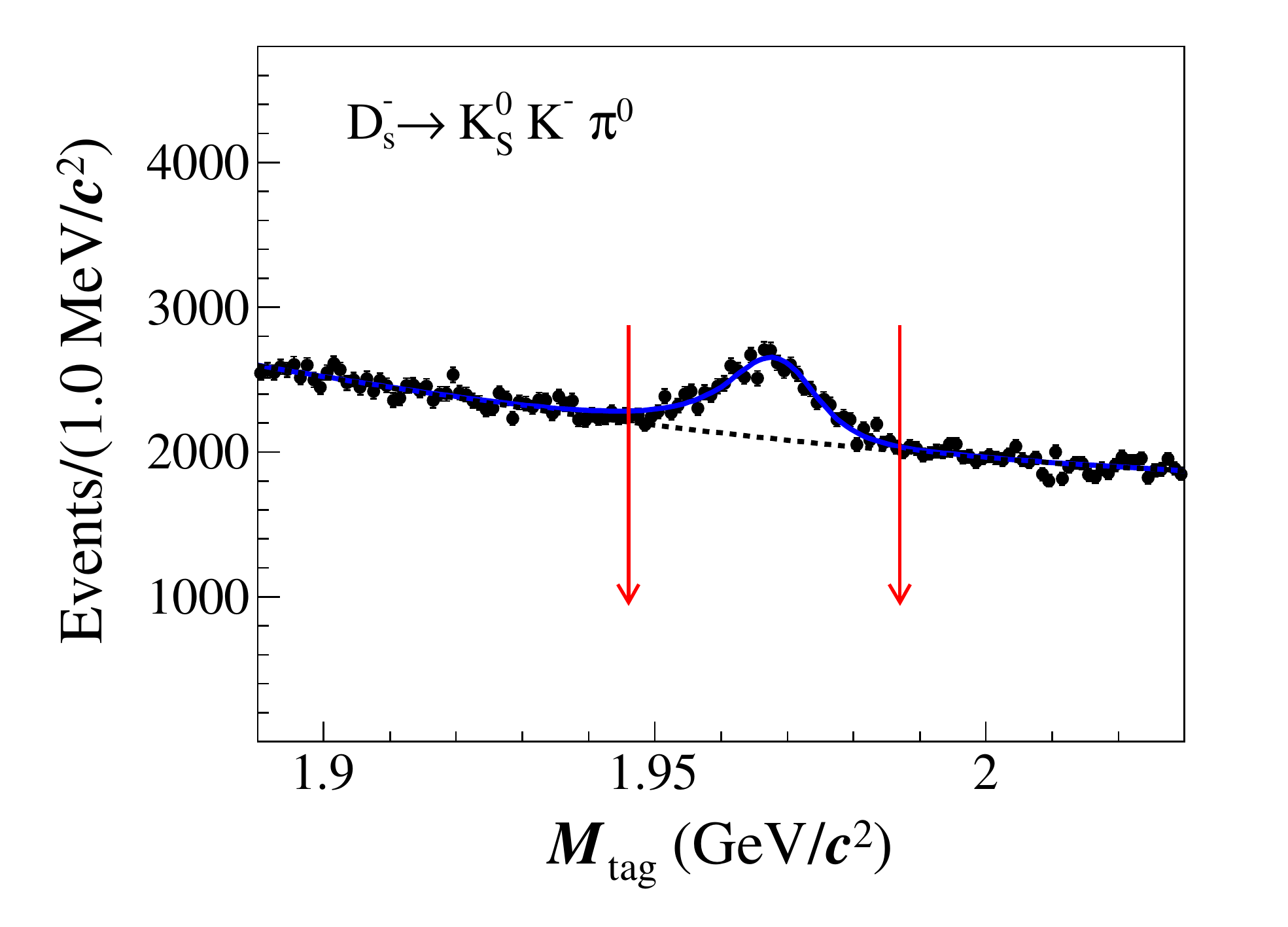}
    \includegraphics[width=6.0cm]{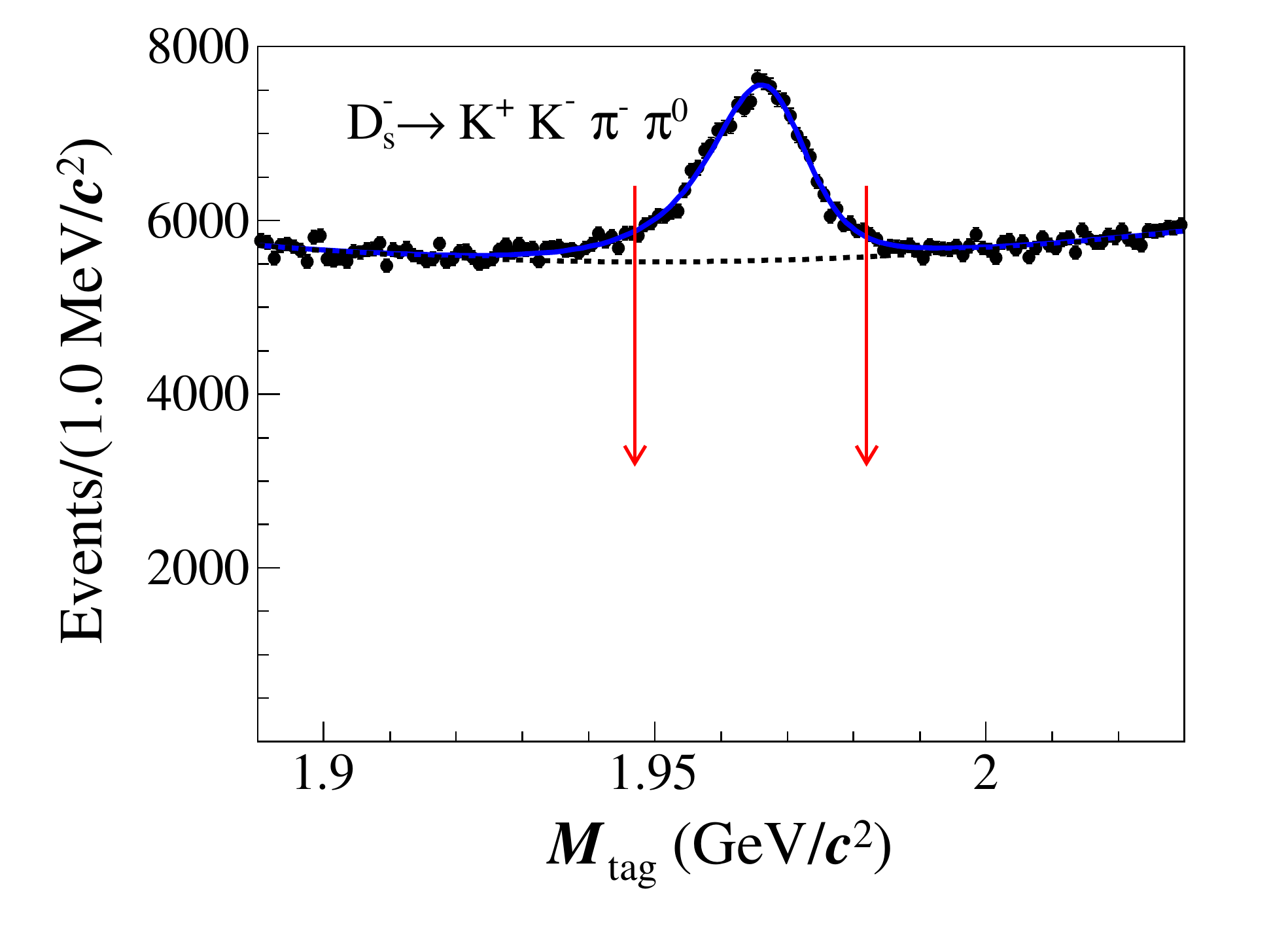}
    \includegraphics[width=6.0cm]{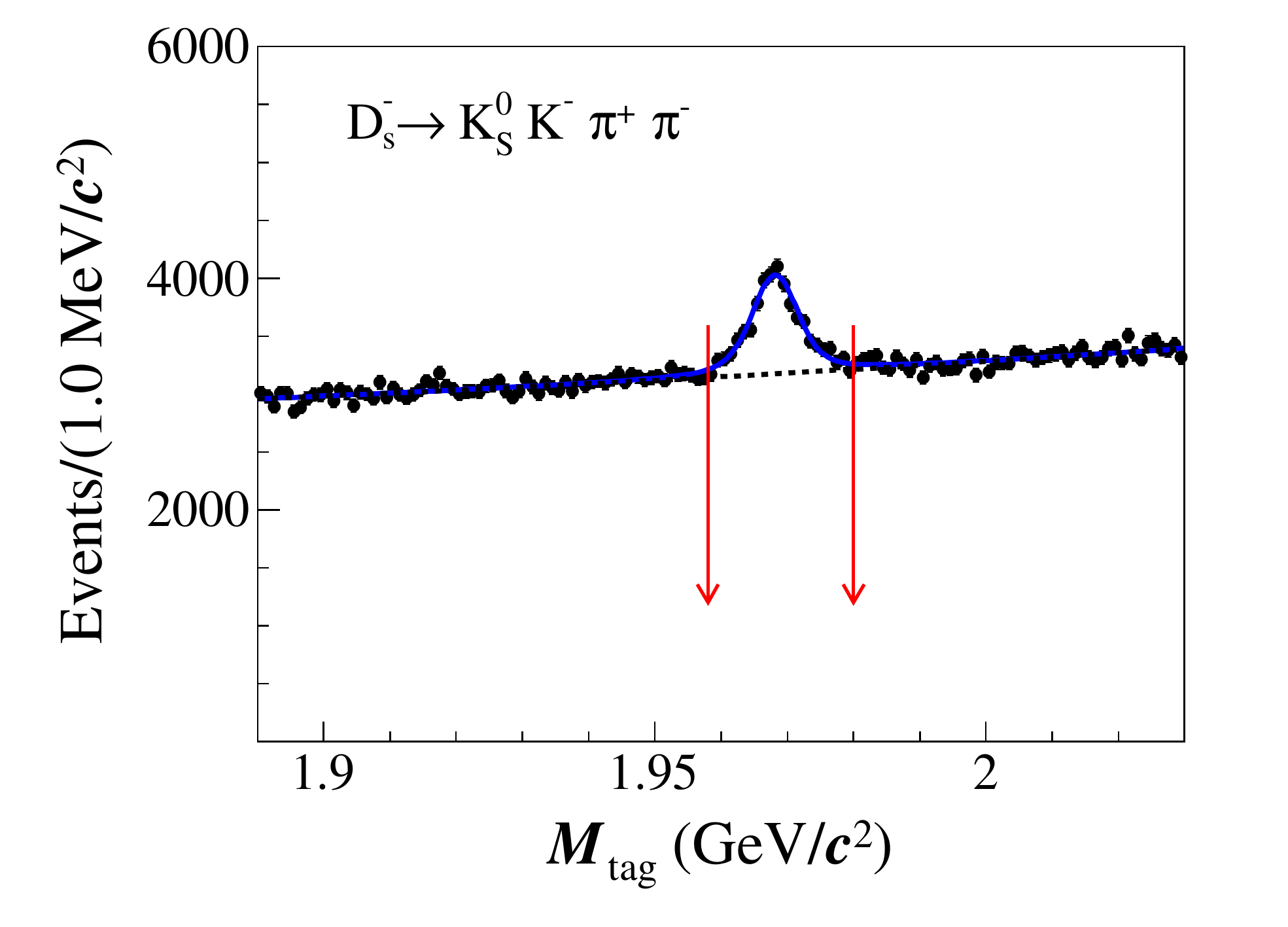}
    \includegraphics[width=6.0cm]{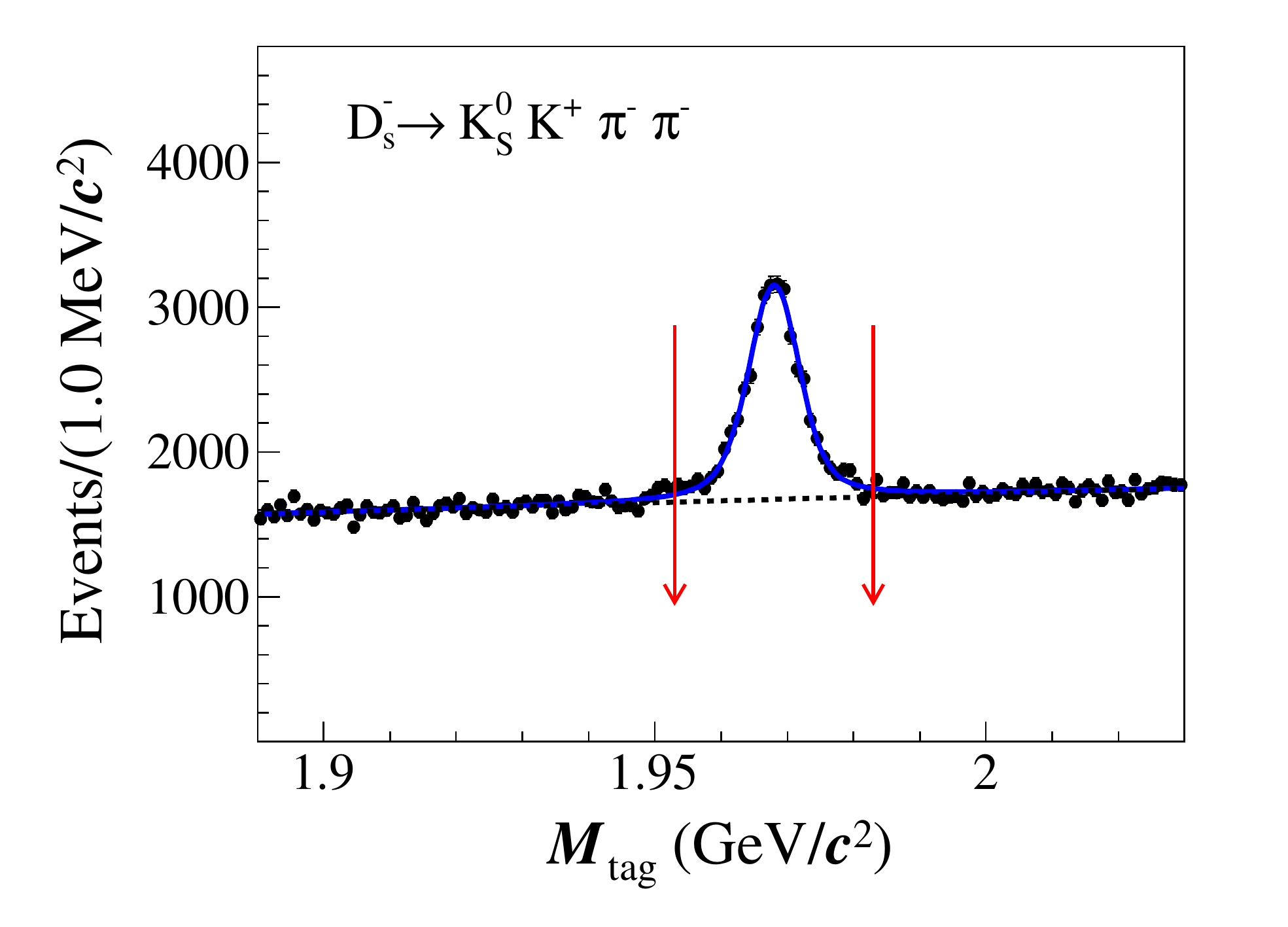}
    \includegraphics[width=6.0cm]{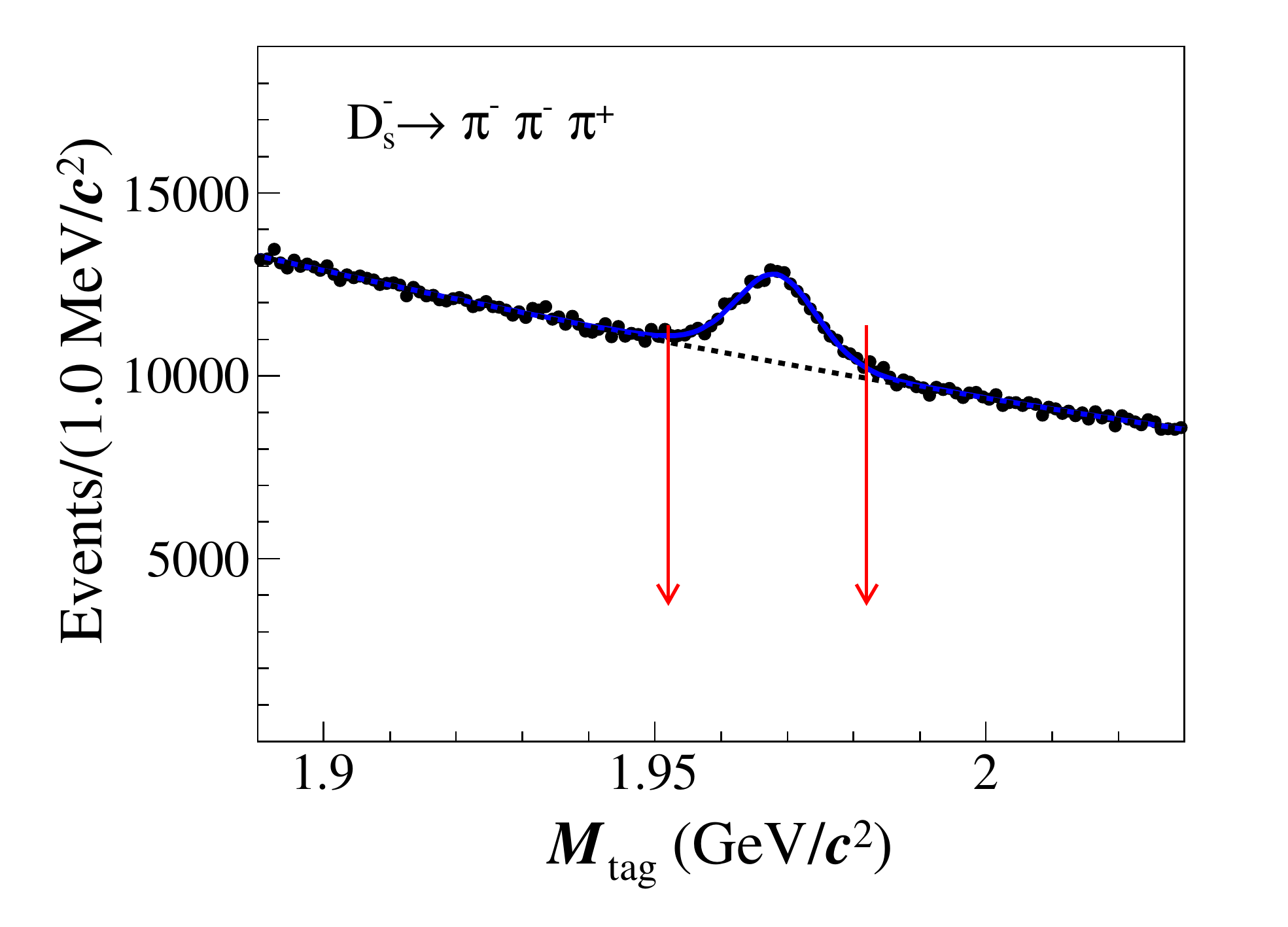}
    \includegraphics[width=6.0cm]{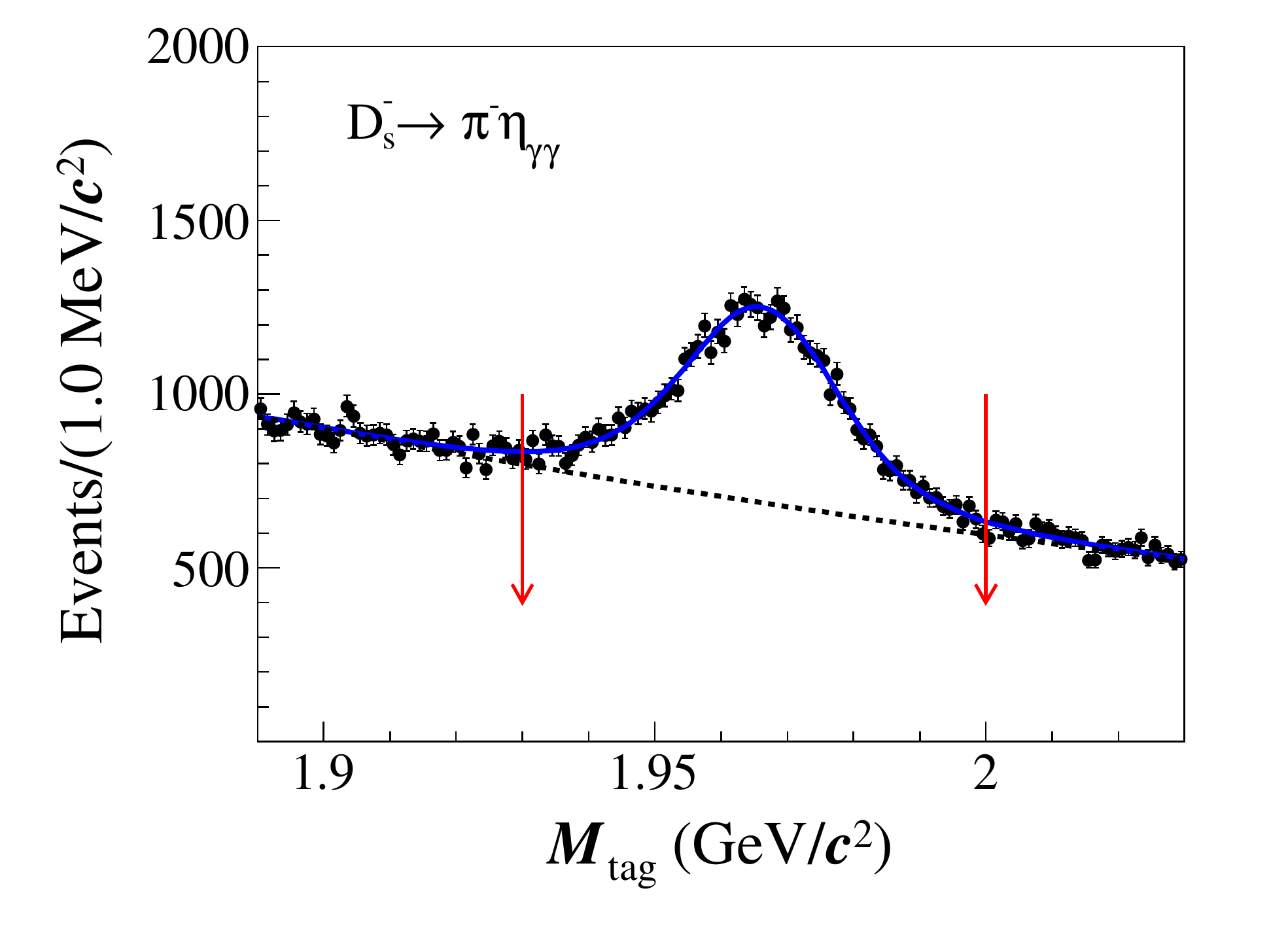}
    \includegraphics[width=6.0cm]{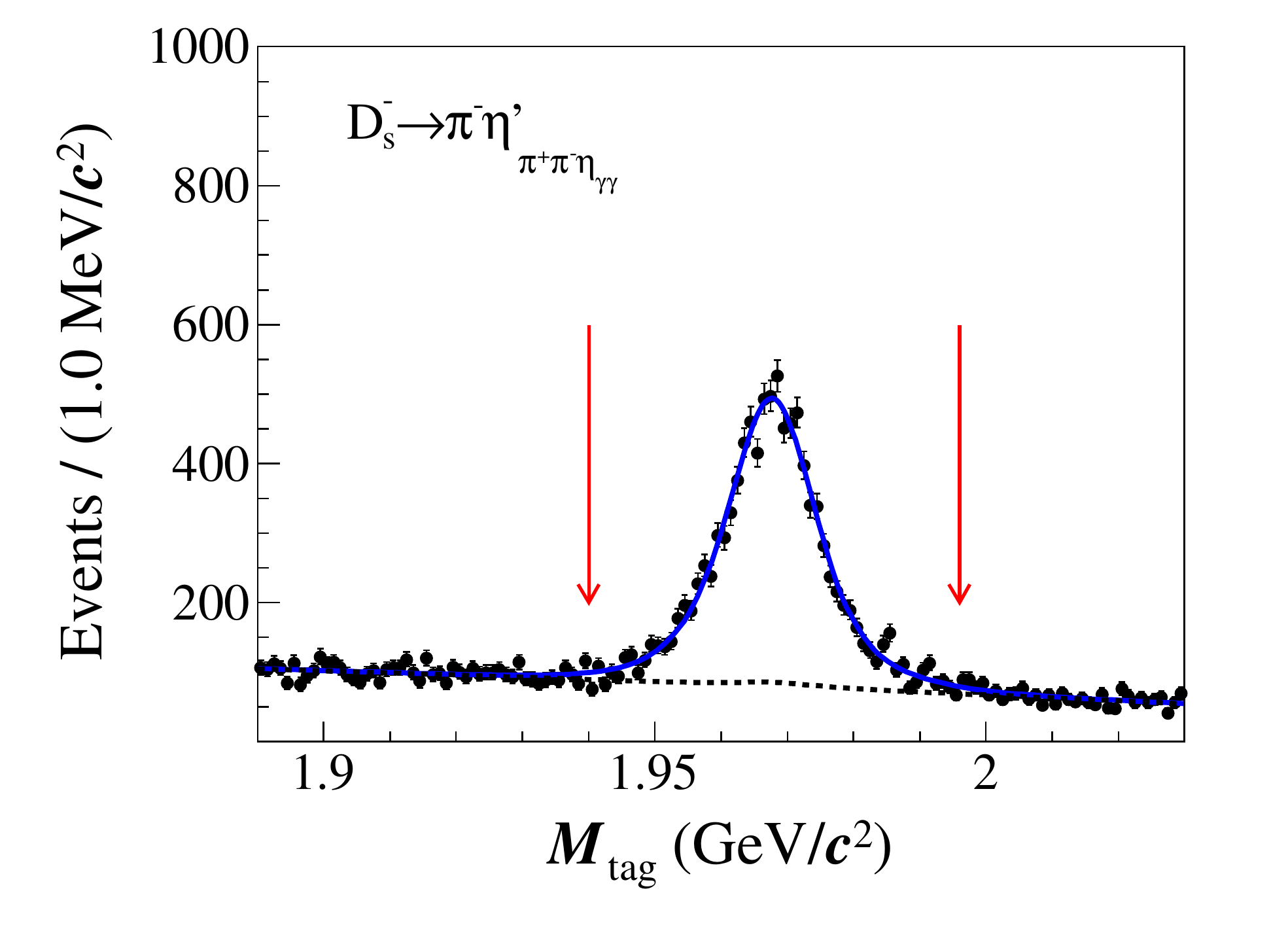}
    \includegraphics[width=6.0cm]{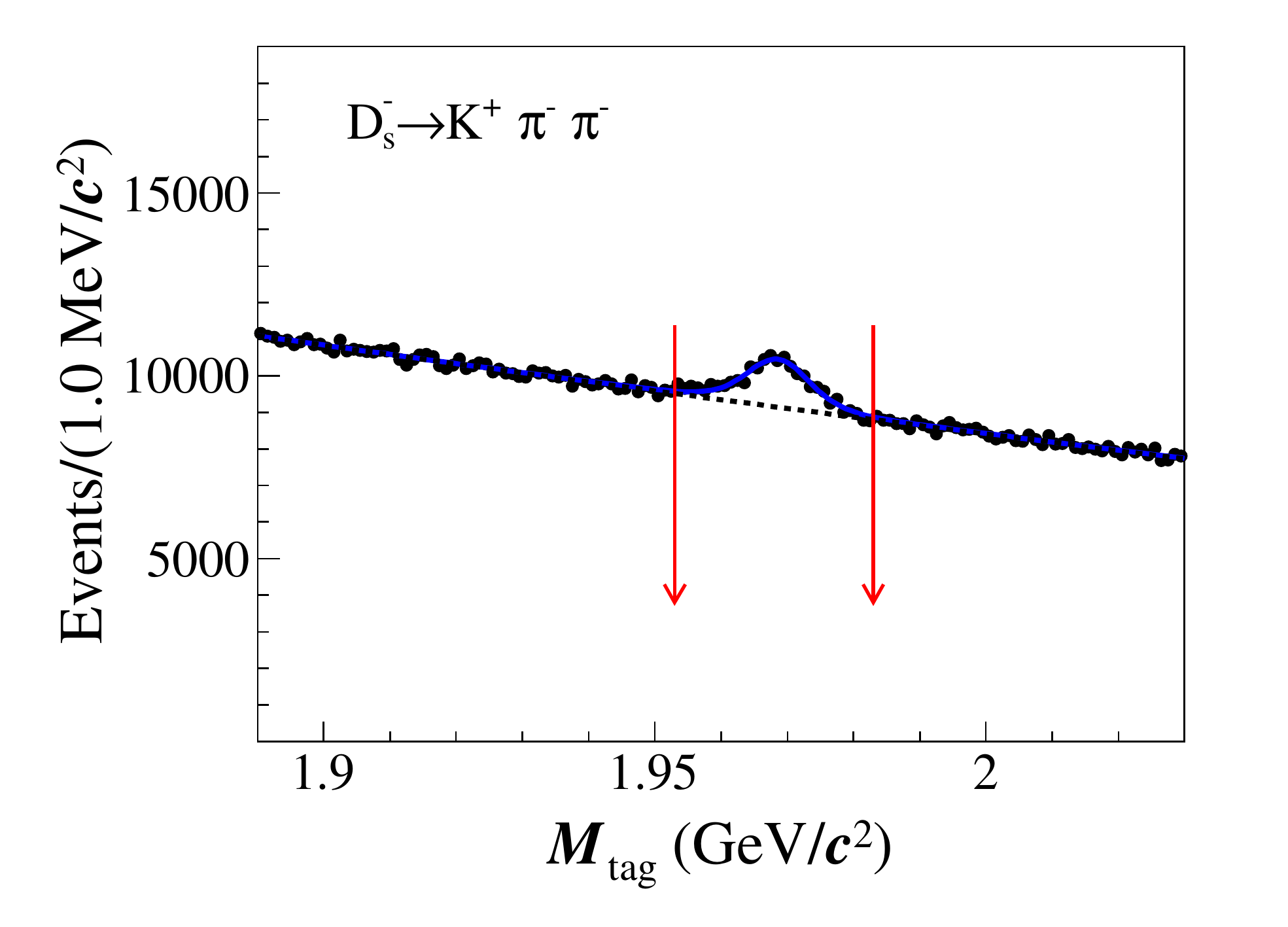}
\caption{Fits to the $M_{\rm tag}$ distributions of the ST candidates from the data sample at $\sqrt{s}=4.178$~GeV. 
	The points with error bars are data, the blue solid lines are the total fits, and the black dashed lines are background. 
	The pairs of red arrows denote the signal regions.}
\label{fit:Mass-data-Ds_4180}
\end{center}
\end{figure*}

In addition to the selection criteria for final-state particles described in Sec.~\ref{ST-selection}, for the branching fraction measurement all the pions are subjected to an additional  momentum cut $p(\pi) > 100$ MeV$/c$, to remove soft pions from $D^{*+}$ decays.
For multiple ST candidates, the candidate with $M_{\rm rec}$ closest to the known mass of $D_s^{*+}$~\cite{PDG} is chosen as the best candidate. 
Besides the tag modes shown in Table~\ref{tagwindow} in the amplitude analysis, we add another two tag modes: $D_s^-\to \pi^-\pi^-\pi^+$ and $D_s^-\to K^-\pi^-\pi^+$. The $M_{\rm tag}$ windows are [1.952, 1.982] and [1.953, 1.986] GeV/$c^2$, respectively.
The yields for various tag modes are listed in Table~\ref{ST-eff}, and they are obtained by fitting the corresponding $M_{\rm tag}$ distributions. 
To prevent an event being double counted in the $D_s^-\to K_S^0K^-$ and $D_s^-\to K^-\pi^{+}\pi^{-}$ selections, the value of $M_{\pi^{+}\pi^{-}}$ is required to be outside of the mass range $[0.487, 0.511]$ GeV$/c^{2}$ for the $D_s^-\to K^-\pi^{+}\pi^{-}$ decay.
As an example, the fits to the data sample at $\sqrt s=4.178$~GeV are shown in Fig.~\ref{fit:Mass-data-Ds_4180}.
In the fits, the signal is modeled by a MC-simulated shape convolved with a Gaussian function to take into account the data-MC difference. 
The background is described by a second-order Chebyshev polynomial. 
For the tag mode $D_{s}^{-} \to K_{S}^{0} K^-$, there are some peaking backgrounds coming from $D^{-} \to K_{S}^{0} \pi^-$. 
The shape of this background is taken from the inclusive MC samples and added to the fit leaving its yield floating. 
For the tag mode $D_{s}^{-} \to \pi^-\eta^{\prime}$, there is the peaking background coming from $D_{s}^{-} \to \eta\pi^+\pi^-\pi^-$. 
The shape and yield of this background are taken from the inclusive MC samples and added to the fit. 
\begin{table*}[htbp]
  \caption{The ST yields for the samples collected at $\sqrt{s} =$ (I) 4.178~GeV, (II) 4.199-4.219~GeV,
    and (III) 4.226~GeV. The uncertainties are statistical.}\label{ST-eff}
    \begin{center}
      \begin{tabular}{lccc}
        \hline
        Tag mode                                    & (I) $N_{\rm ST}$           & (II) $N_{\rm ST}$        & (III) $N_{\rm ST}$      \\
        \hline
        $D_{s}^{-}\to K_{S}^{0}K^{-}$               & $\phantom{0}31941\pm312$   & $18559\pm261$            & $\phantom{0}6582\pm160$ \\
        $D_{s}^{-}\to K^{+}K^{-}\pi^{-}$            & $137240\pm614$             & $81286\pm505$            & $28439\pm327$           \\
        $D_{s}^{-}\to K_{S}^{0}K^{-}\pi^{0}$        & $\phantom{0}11385\pm529$   & $\phantom{0}6832\pm457$  & $\phantom{0}2227\pm220$ \\
				$D^{-}_{s}\to K^{+}K^{-}\pi^{-}\pi^{0}$     & $\phantom{0}39306\pm799$   & $23311\pm659$            & $\phantom{0}7785\pm453$ \\
        $D_{s}^{-}\to K_{S}^{0}K^{-}\pi^{-}\pi^{+}$ & $\phantom{00}8093\pm326$   & $\phantom{0}5269\pm282$  & $\phantom{0}1662\pm217$ \\
        $D_{s}^{-}\to K_{S}^{0}K^{+}\pi^{-}\pi^{-}$ & $\phantom{0}15719\pm289$   & $\phantom{0}8948\pm231$  & $\phantom{0}3263\pm172$ \\
				$D^{-}_{s}\to \pi^{-}\eta_{\gamma\gamma}$   & $\phantom{0}17940\pm402$   & $\phantom{0}10025\pm339$ & $\phantom{0}3725\pm252$ \\
				$D^{-}_{s}\to \pi^{-}\pi^{-}\pi^{+}$        & $\phantom{0}37977\pm859$   & $21909\pm776$            & $\phantom{0}7511\pm393$ \\
        $D_{s}^{-}\to \pi^{-}\eta^{\prime}$         & $\phantom{00}7759\pm141$   & $\phantom{0}4428\pm111$  & $\phantom{0}1648\pm74\phantom{0}$ \\
        $D_{s}^{-}\to K^{-}\pi^{+}\pi^{-}$          & $\phantom{0}17423\pm666$   & $10175\pm448$            & $\phantom{0}4984\pm458$ \\
        \hline
      \end{tabular}
    \end{center}
\end{table*}

Once a tag mode is identified, we search for the signal decay $D_{s}^{+} \to K^{-}K^{+}\pi^{+}\pi^{+}\pi^{-}$ at the recoiling side. 
In the case of multiple candidates, the DT candidate with the average mass, $(M_{\rm sig}+M_{\rm tag})/2$, closest to the $D_{s}^{\pm}$ nominal mass is retained.

To measure the BF, we start from the following equations for one ST mode:
\begin{eqnarray}\begin{aligned}
  N_{\text{tag}}^{\text{ST}} = 2N_{D_{s}^{+}D_{s}^{-}}\mathcal{B}_{\text{tag}}\epsilon_{\text{tag}}^{\text{ST}}\,, \label{eq-ST}
\end{aligned}\end{eqnarray}
\begin{equation}
  N_{\text{tag,sig}}^{\text{DT}}=2N_{D_{s}^{+}D_{s}^{-}}\mathcal{B}_{\text{tag}}\mathcal{B}_{\text{sig}}\epsilon_{\text{tag,sig}}^{\text{DT}}\,,
  \label{eq-DT}
\end{equation}
where $N_{\text{tag}}^{\text{ST}}$ is the ST yield for the tag mode, $N_{\text{tag,sig}}^{\text{DT}}$ is the DT yield, 
$N_{D_{s}^{+}D_{s}^{-}}$ is the total number of $D_{s}^{*\pm}D_{s}^{\mp}$ pairs produced in the $e^{+}e^{-}$ collisions, 
$\mathcal{B}_{\text{tag}}$ and $\mathcal{B}_{\text{sig}}$ are the BFs of the tag and signal modes, respectively, $\epsilon_{\text{tag}}^{\text{ST}}$ is the ST efficiency to reconstruct the tag mode and $\epsilon_{\text{tag,sig}}^{\text{DT}}$ is the DT efficiency to reconstruct both the tag and signal decay modes. 
In the case of more than one tag mode and sample group,
\begin{eqnarray}
\begin{aligned}
  \begin{array}{lr}
    N_{\text{total}}^{\text{DT}}=\Sigma_{\alpha, i}N_{\alpha,\text{sig},i}^{\text{DT}}   = \mathcal{B}_{\text{sig}}
		\Sigma_{\alpha, i}2N^{i}_{D_{s}^{+}D_{s}^{-}}\mathcal{B}_{\alpha}\epsilon_{\alpha,\text{sig}, i}^{\text{DT}}\,,
  \end{array}
  \label{eq-DTtotal}
\end{aligned}
\end{eqnarray}
where $\alpha$ represents the tag modes in the $i^{\rm th}$ sample group. 
We isolate $\mathcal{B}_{\text{sig}}$ by using Eq.~(\ref{eq-ST}):
\begin{eqnarray}\begin{aligned}
  \mathcal{B}_{\text{sig}} =
  \frac{N_{\text{total}}^{\text{DT}}}{ \begin{matrix}\sum_{\alpha, i} N_{\alpha, i}^{\text{ST}}\epsilon^{\text{DT}}_{\alpha,\text{sig},i}/\epsilon_{\alpha,i}^{\text{ST}}\end{matrix}},
\end{aligned}\end{eqnarray}
where $N_{\alpha,i}^{\text{ST}}$ and $\epsilon_{\alpha,i}^{\text{ST}}$ are obtained from the data and inclusive MC samples, respectively, and $\epsilon_{\alpha,\text{sig},i}^{\text{DT}}$ is determined with signal MC samples. 
The decay of $D_{s}^{+} \to K^{-}K^{+}\pi^{+}\pi^{+}\pi^{-}$ events is generated according to the results of the amplitude analysis.  

\begin{figure}[!htbp]
  \centering
  \includegraphics[width=0.6\textwidth]{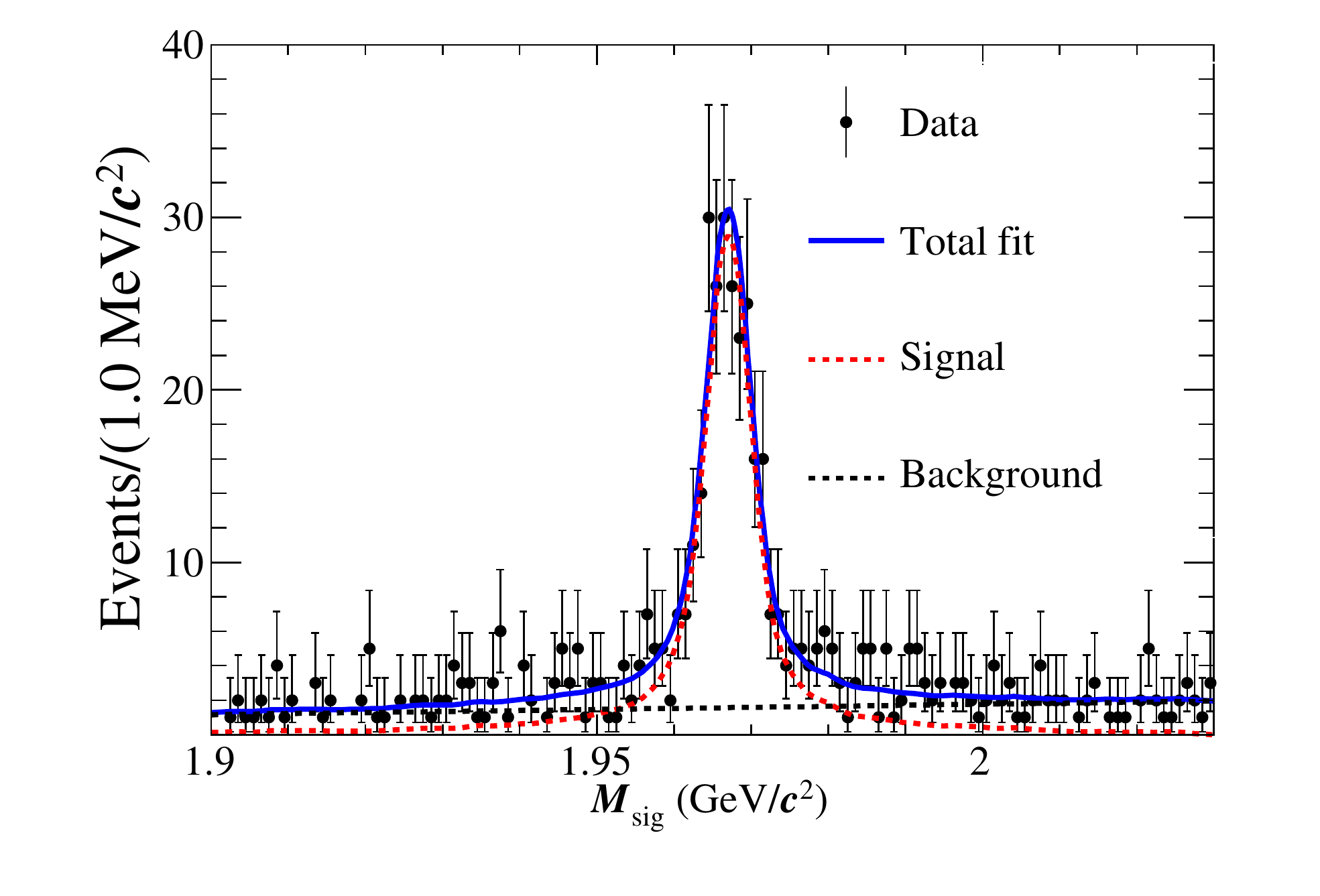}
 \caption{Fit to the $M_{\rm sig}$ distribution of the DT candidates from the combined data samples at $\sqrt{s}= 4.178$-$4.226$~GeV. 
	The data are represented by points with error bars, the total fit by the blue solid line, and the fitted signal and the fitted background by the red dotted and the black dashed lines, respectively.}
  \label{DT-fit}
\end{figure}

The DT yield $N_{\text{total}}^{\text{DT}}$ is found to be $309\pm22$ from the fit to the $M_{\rm sig}$ distribution of the selected DT candidates, with purity $(60.4\pm2.8)\%$ for the data samples at $\sqrt{s}= 4.178$-$4.226$~GeV. 
The fit result is shown in Fig.~\ref{DT-fit}, where the signal shape is described by a MC-simulated shape convolved with a Gaussian function and the background shape is described by a linear function.
Taking into account the differences in $K^{\pm}$ and $\pi^{\pm}$ tracking and PID efficiencies between data and MC simulation, we determine the BF of $D^+_s\to K^-K^+\pi^+\pi^+\pi^-$ to be $(6.60\pm0.47_{\rm stat.}\pm0.35_{\rm syst.})\times 10^{-3}$.

The uncertainty in the total yield of the ST $D^{-}_{s}$ mesons is assigned to be 0.4$\%$ by taking into account the background fluctuations in the fit, and by examining the changes of the fit yields when varying the signal and the background shapes. 
The tracking and PID efficiencies of $K^{\pm}$ are studied with $e^+e^-\to K^+K^-K^+K^-$ and $e^+e^- \to K^+K^-\pi^+\pi^-(\pi^0)$ events. 
The data-MC efficiency ratios of $K^{+}$ ($K^{-}$) tracking and PID efficiencies, weighted by the corresponding momentum spectra from signal MC events, are $1.005\pm0.017$ ($0.998\pm0.015$) and $0.983\pm0.003$ ($0.983\pm0.003$), respectively. 
After correcting the MC efficiencies for these averaged data-MC differences, the systematic uncertainties of tracking and PID efficiencies per $K^{+}$ ($K^{-}$) are assigned as $1.7\%$ ($1.5\%$) and $0.3\%$ ($0.3\%$), respectively.
The $\pi^{\pm}$ tracking and PID efficiencies are studied with $e^+e^- \to K^+K^-\pi^+\pi^-$ events.
The data-MC efficiency ratios of the $\pi^{+}$ ($\pi^{-}$) tracking and PID efficiencies are $0.999\pm0.005$ ($0.990\pm0.005$) and $1.004\pm0.002$ ($1.004\pm0.002$), respectively,
and we assign $0.5\%$ ($0.5\%$) and $0.2\%$ ($0.2\%$) as the systematic uncertainties arising from $\pi^{\pm}$ tracking and PID, respectively.

The systematic uncertainty due to the signal shape is studied by repeating the fit without the convolved Gaussian function. 
For the background shape of the signal $D^{+}_{s}$, the MC-simulated shape is used to replace the linear function.
The difference of the DT yields is taken as the systematic uncertainty.
The uncertainty due to the limited MC statistics is obtained by $\sqrt{\begin{matrix} \sum_{i} (f_{i}\frac{\delta_{\epsilon_{i}}}{\epsilon_{i}}\end{matrix}})^2$,
where $f_{i}$ is the tag yield fraction, and $\epsilon_{i}$ and $\delta_{\epsilon_{i}}$ are the signal efficiency and the corresponding uncertainty of tag mode $i$, respectively.
The uncertainty from the amplitude model is estimated by varying the model parameters based on their error matrix. 
The distribution of 300 efficiency values resulting from this variation are fitted by a Gaussian function and the fitted resolution divided by the mean value is taken as uncertainty.
All of the systematic uncertainties are summarised in Table~\ref{BF-Sys}.
Adding them in quadrature gives a total systematic uncertainty in the BF measurement of 5.4\%.
\begin{table}[htbp]
  \caption{Systematic uncertainties in the BF measurement.}
  \label{BF-Sys}
  \begin{center}
    \begin{tabular}{lccc}
      \hline
      Source   & Systematic uncertainty (\%)\\
      \hline
		ST yield   &0.4\\
    Tracking         &4.2\\
    PID              &1.0\\
    Signal shape     &0.6\\
    Background shape &1.9\\
    MC statistics    &0.6\\
    PWA model        &2.4\\
    \hline
    Total            &5.4\\
      \hline
    \end{tabular}
  \end{center}
\end{table}

\section{Summary}
An amplitude analysis of the decay $D^+_s\to K^-K^+\pi^+\pi^+\pi^-$ has been performed for the first time. 
Amplitudes with statistical significances larger than 5$\sigma$ are selected. 
The results for the FFs and phases of the different intermediate processes are listed in Table~\ref{tab:tested_amplitudes}. 
With the detection efficiency determined according to the results from the amplitude analysis, the BF for the decay $D^+_s\to K^-K^+\pi^+\pi^+\pi^-$ is measured to be $(6.60\pm0.47_{\rm stat.}\pm0.35_{\rm syst.})\times 10^{-3}$, 
which is consistent with the PDG value of $(8.6\pm1.5)\times 10^{-3}$ within $1.5\sigma$.
The BFs of the intermediate processes calculated with $\mathcal{B}_i = \rm FF$$_i \times \mathcal{B}$$(D^+_s\to K^-K^+\pi^+\pi^+\pi^-)$ in this analysis and those world average values from the PDG~\cite{PDG} are listed in Table~\ref{inter-processes}.
The precision is improved by about a factor of two compared to the world average value. 
\begin{table}[htbp]
  \caption{The BFs for various intermediate processes measured in this analysis and from the PDG~\cite{PDG}, the first and second uncertainties are statistical and systematic, respectively.}
  \centering
  \begin{tabular}{lcc}
  \hline
      \hline
		Intermediate process  &BF ($10^{-3}$)  &PDG ($10^{-3}$)\\
  \hline
		$D^+_s[S]\to a_1(1260)^+\phi, a_1(1260)^+[S]\to\rho^0\pi^+$ &$4.8\pm0.4\pm0.3$ &\\
    $D^+_s[P]\to a_1(1260)^+\phi, a_1(1260)^+[S]\to\rho^0\pi^+$ &$0.3\pm0.1\pm0.1$ &\\
    $D^+_s\to a_1(1260)^+\phi$ &$5.2\pm0.4\pm0.3$ &$7.4\pm1.2$\\
    $D^+_s\to (K^-K^+\pi^+\pi^+\pi^-)_{\rm NR}$                     &$1.4\pm0.2\pm0.1$ &$0.9\pm0.7$\\
        \hline
      \hline
  \end{tabular}
  \label{inter-processes}
\end{table}

\acknowledgments
The BESIII collaboration thanks the staff of BEPCII and the IHEP computing center for their strong support. This work is supported in part by National Key Research and Development Program of China under Contracts Nos. 2020YFA0406400, 2020YFA0406300; National Natural Science Foundation of China (NSFC) under Contracts Nos. 11625523, 11635010, 11735014, 11775027, 11822506, 11835012, 11875054, 11935015, 11935016, 11935018, 11961141012, 12192260, 12192261, 12192262, 12192263, 12192264, 1219226; the Chinese Academy of Sciences (CAS) Large-Scale Scientific Facility Program; Joint Large-Scale Scientific Facility Funds of the NSFC and CAS under Contracts Nos. U1732263, U1832207, U1932108, U2032104; CAS Key Research Program of Frontier Sciences under Contracts Nos. QYZDJ-SSW-SLH003, QYZDJ-SSW-SLH040; 100 Talents Program of CAS; INPAC and Shanghai Key Laboratory for Particle Physics and Cosmology; ERC under Contract No. 758462; European Union Horizon 2020 research and innovation programme under Contract No. Marie Sklodowska-Curie grant agreement No 894790; German Research Foundation DFG under Contracts Nos. 443159800, Collaborative Research Center CRC 1044, FOR 2359, FOR 2359, GRK 214; Istituto Nazionale di Fisica Nucleare, Italy; Ministry of Development of Turkey under Contract No. DPT2006K-120470; National Science and Technology fund; Olle Engkvist Foundation under Contract No. 200-0605; STFC (United Kingdom); The Knut and Alice Wallenberg Foundation (Sweden) under Contract No. 2016.0157; The Royal Society, UK under Contracts Nos. DH140054, DH160214; The Swedish Research Council; U. S. Department of Energy under Contracts Nos. DE-FG02-05ER41374, DE-SC-0012069.

\bibliographystyle{JHEP}
\bibliography{references}

\clearpage
\appendix
\section{All other tested amplitudes}
\label{text}
\begin{sloppypar}
All other tested amplitudes' significances are less than 5 $\sigma$, so they are not included in nominal fit and their significances are listed in Table~\ref{tab:tested_amplitude}.
\end{sloppypar}

\begin{table}[htbp]
  \caption{Tested amplitudes, but not included in the nominal fit.}
  \label{tab:tested_amplitude}
  \begin{center}
    \footnotesize
    \begin{tabular}{lc}
      \toprule\toprule
      \hline
      Tested Amplitude                                          &Significance($\sigma$)\\
      \hline
      $D_s^{+}[S]\to a_1(1260)^{+}\phi, a_1(1260)^{+}[D]\to \rho\pi^{+}, \phi\to K^-K^+$ &2.7\\
      $D_s^{+}[P]\to a_1(1260)^{+}\phi, a_1(1260)^{+}[D]\to \rho\pi^{+}, \phi\to K^-K^+$ &$<1$\\
      $D_s^{+}[D]\to a_1(1260)^{+}\phi, a_1(1260)^{+}[S]\to \rho\pi^{+}, \phi\to K^-K^+$ &1.5\\
      $D_s^{+}[D]\to a_1(1260)^{+}\phi, a_1(1260)^{+}[D]\to \rho\pi^{+}, \phi\to K^-K^+$ &1.5\\
      $D_s^{+}[S]\to a_1(1260)^{+}\phi, a_1(1260)^{+}[P]\to f_0(500)\pi^{+}, \phi\to K^-K^+$ &$<1$\\
      $D_s^{+}[P]\to a_1(1260)^{+}\phi, a_1(1260)^{+}[P]\to f_0(500)\pi^{+}, \phi\to K^-K^+$ &$<1$\\
      $D_s^{+}[D]\to a_1(1260)^{+}\phi, a_1(1260)^{+}[P]\to f_0(500)\pi^{+}, \phi\to K^-K^+$ &1.5\\
      $D_s^+[P]\to a_1(1260)^+a_0(980), a_1(1260)^+[S]\to\rho^0\pi^+$, $a_0(980)\to K^-K^+$ &$<1$\\
      $D_s^+[P]\to a_1(1260)^+a_0(980), a_1(1260)^+[D]\to\rho^0\pi^+$, $a_0(980)\to K^-K^+$ &3.6\\
      $D_s^+[S]\to K_1(1270)^{+}\bar{K^{*0}(892)}, K_1(1270)[S]\to \rho K^{+}, \bar{K^{*0}(892)}\to K^-\pi^+$ &$<1$\\
      $D_s^+[S]\to K_1(1270)^{+}\bar{K^{*0}(892)}, K_1(1270)[D]\to \rho K^{+}, \bar{K^{*0}(892)}\to K^-\pi^+$ &$<1$\\
      $D_s^+[P]\to K_1(1270)^{+}\bar{K^{*0}(892)}, K_1(1270)[S]\to \rho K^{+}, \bar{K^{*0}(892)}\to K^-\pi^+$ &$<1$\\
      $D_s^+[P]\to K_1(1270)^{+}\bar{K^{*0}(892)}, K_1(1270)[D]\to \rho K^{+}, \bar{K^{*0}(892)}\to K^-\pi^+$ &$<1$\\
      $D_s^+[D]\to K_1(1270)^{+}\bar{K^{*0}(892)}, K_1(1270)[S]\to \rho K^{+}, \bar{K^{*0}(892)}\to K^-\pi^+$ &1.5\\
      $D_s^+[D]\to K_1(1270)^{+}\bar{K^{*0}(892)}, K_1(1270)[D]\to \rho K^{+}, \bar{K^{*0}(892)}\to K^-\pi^+$ &1.5\\
      $D_s^{+}[S]\to \pi^{+}\pi^{+}\pi^{-}\phi, \phi\to K^-K^+$ &$<1$\\
      \hline
      \bottomrule\bottomrule
    \end{tabular}
  \end{center}
\end{table}

\large
The BESIII Collaboration\\
\normalsize
\\M.~Ablikim$^{1}$, M.~N.~Achasov$^{10,b}$, P.~Adlarson$^{67}$, S. ~Ahmed$^{15}$, M.~Albrecht$^{4}$, R.~Aliberti$^{28}$, A.~Amoroso$^{66A,66C}$, M.~R.~An$^{32}$, Q.~An$^{63,49}$, X.~H.~Bai$^{57}$
% ...
\\
\vspace{0.2cm} {\it
$^{1}$ Institute of High Energy Physics, Beijing 100049, People's Republic of China\\
% ...
$^{4}$ Bochum Ruhr-University, D-44780 Bochum, Germany\\
% ...
$^{10}$ G.I. Budker Institute of Nuclear Physics SB RAS (BINP), Novosibirsk 630090, Russia\\
% ...
$^{15}$ Helmholtz Institute Mainz, Staudinger Weg 18, D-55099 Mainz, Germany\\
% ...
$^{28}$ Johannes Gutenberg University of Mainz, Johann-Joachim-Becher-Weg 45, D-55099 Mainz, Germany\\
% ...
$^{32}$ Liaoning Normal University, Dalian 116029, People's Republic of China\\
% ...
$^{49}$ State Key Laboratory of Particle Detection and Electronics, Beijing 100049, Hefei 230026, People's Republic of China\\
% ...
$^{57}$ University of Jinan, Jinan 250022, People's Republic of China\\
% ...
$^{63}$ University of Science and Technology of China, Hefei 230026, People's Republic of China\\
% ...
$^{66}$ University of Turin and INFN, (A)University of Turin, I-10125, Turin, Italy; (B)University of Eastern Piedmont, I-15121, Alessandria, Italy; (C)INFN, I-10125, Turin, Italy\\
$^{67}$ Uppsala University, Box 516, SE-75120 Uppsala, Sweden\\
% ...
$^{b}$ Also at the Novosibirsk State University, Novosibirsk, 630090, Russia\\
% ...
}

\end{document}